\documentclass[12pt,review]{elsarticle}
\usepackage[margin=1in]{geometry}
\usepackage{amsmath}
\usepackage{amssymb}
\usepackage{bbm}
\usepackage{bm}
\usepackage{natbib}
\biboptions{authoryear}
\usepackage{blkarray, bigstrut}
\usepackage{setspace}
\usepackage{url}
\usepackage{makecell}
\usepackage{dirtytalk}
\usepackage{subfig}

\newtheorem{Def}{Definition}
\newtheorem{Remark}[Def]{Remark}

\newtheorem{Lemma}[Def]{Lemma}

\newtheorem{Proposition}[Def]{Proposition}
\newcommand{\proof}{\noindent \textbf{Proof:}  \ }

\makeatletter
\def\ps@pprintTitle{%
 \let\@oddhead\@empty
 \let\@evenhead\@empty
 \def\@oddfoot{\hfill}%\today}%
 \let\@evenfoot\@oddfoot}
\makeatother

\begin{document}

\begin{frontmatter}



\title{Confidence sets for dynamic poverty indexes}
\author[1]{Guglielmo D'Amico\corref{cor1}} \ead{g.damico@unich.it}
\author[2]{Riccardo De Blasis}
\author[3]{Philippe Regnault}
\address[1]{Universit\`a "G. D’Annunzio", Chieti, Italy}
\address[2]{Universit\`a Politecnica delle Marche, Ancona, Italy}
\address[3]{Universit\'e de Reims Champagne-Ardenne, Reims, France}

\cortext[cor1]{Corresponding author}


\begin{abstract}
    In this study, we extend the research on the dynamic poverty indexes, namely the dynamic Headcount ratio, the dynamic income-gap ratio, the dynamic Gini and the dynamic Sen, proposed in \cite{damico2018DynamicMeasurementPoverty}. The contribution is twofold. First, we extend the computation of the dynamic Gini index, thus the Sen index accordingly, with the inclusion of the inequality within each class of poverty where people are classified according to their income. Second, for each poverty index, we establish a central limit theorem that gives us the possibility to determine the confidence sets. An application to the Italian income data from 1998 to 2012 confirms the effectiveness of the considered approach and the possibility to determine the evolution of poverty and inequality in real economies.
\end{abstract}


\end{frontmatter}
	
\section{Introduction}
Economists and econometricians have dedicated a lot of efforts to the investigation of poverty. The literature is vast and comprehends theoretically oriented contributions as well as applied researches. The advancement to more powerful indexes of poverty is always of interest and it aims at capturing specific peculiarities of the phenomenon that were ignored by previous indicators \citep[see, e.g.,][]{hirsch2020LowIncomeGap}. In particular, multidimensional measures of poverty are relevant in this context, with the inclusion of non-monetary sources of deprivation which affect the well-being of individuals and households such as disability, exposure to environmental hazards and limited availability of healthcare services \citep[see, e.g.,][] {park2020MultidimensionalPovertyStatus,annoni2015MultidimensionalViewPoverty,parodi2008DisabilityItalianHouseholds}. This research stream has his origin in the seventies of the last century when the stately contribution by Professor Sen appeared  \citep{sen1976PovertyOrdinalApproach}. Since then, continuous improvements and generalisations have been made \citep[see, e.g.,][]{takayama1979PovertyIncomeInequality,shorrocks1995RevisitingSenPoverty,foster2009ClassChronicPoverty}.

Almost at the same time, it was recognised that poverty is not a static notion and that this characteristic should be investigated in relation to time. The main idea is to statistically assess frequencies of poverty condition and their variations through time. These statistical properties of poverty mobility were determined and confirmed in several studies \citep[see, e.g.,][]{bane1986SlippingOutPoverty,duncan1993PovertyDynamicsEight,whelan2000POVERTYDYNAMICSAnalysis}. In general, understanding the features of poverty over time needs the adoption of a stochastic model of income evolution. Therefore, Markov chain models were frequently used. Some examples are available in  \cite{mccall1971MarkovianModelIncome}, \cite{breen2004PovertyDynamicsCorrected}, \cite{cappellari2004ModellingLowIncome}, \cite{formby2004MobilityMeasurementTransition}, and \cite{langheheine2016UnifyingFrameworkMarkov}. In addition, poverty rates and transition probabilities have been estimated in relation to noisy data by \cite{lee2017EstimationPovertyTransition}.

In contrast to previous research, the idea of advancing dynamic indexes of poverty and income inequality is relatively new and limited to a few contributions. Precisely, at authors knowing, two approaches can be identified. They share the same starting idea, namely the extension of static indicators into a dynamic framework, but then move away in relation to methods and results to answer different questions. On the one hand, \cite{ewald2015IncreasingRiskInequality} consider a sequence of distributions parametrised in time and they look for conditions under which the corresponding sequence of the indicator (in the specific case, the Gini index) increases over time. On the other hand, starting from the work by  \cite{damico2010GeneralizedConcentrationInequality}, \cite{damico2012IncomeInequalityDynamic} and \cite{damico2014DecompositionPopulationDynamic} built up an economic system by advancing a set of assumptions on the time evolution of income for every agent member of the economic system. This approach was also adopted in \cite{damico2018DynamicMeasurementPoverty} in relation to dynamic measures of poverty where the dynamic indexes were evaluated both for finite and infinite size economic system. Specifically, the infinite size system (i.e., an economy with an infinite number of agents) revealed to be particularly interesting. In fact, for each index, using probabilistic arguments (i.e., strong law of large numbers), it is possible to determine a deterministic function (of the parameters of the model) to which the index of the real economy converges to.

In this paper, we move further steps into this direction. First, we extend the computation of the dynamic Gini index including the inequality within each class of poverty where people are classified according to their income. This extension impacts also the Sen index that is a function of the Gini index. Second, for each poverty index, we establish a central limit theorem that gives us the possibility to determine confidence sets, i.e., bounds that at a fixed probability level express the goodness of the approximations based on the strong law of large numbers. These results derive from the specific assumptions defining the model which are based on the probabilistic equality of the incomes of people belonging to the same income class and that may migrate in time from one class to another according to a continuous time Markov process. Finally, we present an application of the aforementioned probabilistic approximations on the Italian income data provided by the Italian Central Bank from 1998 to 2012 which contains information about family net disposable incomes and household members. The results of the application suggests the effectiveness of the considered approach and confirm the possibility to apply it for the determination of the evolution of poverty and inequality in real economies.\\ 
The remainder of the paper is organised as follows. Section 2 sets out the assumptions that define the model and presents the main theoretical results of the paper including probabilistic approximations of the indexes and their confidence sets. Section 3 illustrates the result of the application to real data and demonstrates the adequacy of the proposed approach to the investigation of the evolution in time of dynamic indexes of poverty in real economic systems. Section 4 summarises our contribution and results. All proofs are deferred to the Appendix.

\section{The stochastic model and confidence sets for dynamic poverty indexes}
\label{model}
    
In this section we present the mathematical model. First, we introduce the stochastic model of income evolution and the dynamic version of four poverty indexes in the case of infinite size economic systems. Then, we derive the confidence sets for the poverty indexes. They are obtained by proving the central limit theorem for the stochastic processes expressing the dynamic poverty indexes.

Following \cite{damico2018DynamicMeasurementPoverty}, we consider an economic system composed of a set $\mathcal{H}$ of $N$ individuals. The income produced by each economic agent evolves randomly in time and can be described through a stochastic process ${\bf Y}=(Y_h(t))_{t \in \mathbb{R}_+}$, where $h$ denotes the h-th individual in the economic system and $t$ is the time variable. For our purposes we classify individuals according to their income in one of three exhaustive and exclusive income classes denoted by a random process $C_h(t)$ such that:
\begin{displaymath}
 C_h(t) := \left\{
            \begin{array}{ll}
                \mathcal{C}_1 & \textrm{if } Y_h(t) \leq y_{ep}, \\ 
                \mathcal{C}_2 & \textrm{if } y_{ep} < Y_h(t) \leq y_p, \\
                \mathcal{C}_{3} & \textrm{if } Y_h(t) > y_p,
            \end{array} \right.
\end{displaymath}
where $y_p$ and $y_{ep}$ are the poverty and extreme poverty threshold rates, respectively. Clearly, the possibility to extend the model to multiple richness classes is straightforward. 

In the remainder of the paper, the simplifying notation $\{1,2,3\}$ will be used to denote the set $\{\mathcal{C}_1, \mathcal{C}_2, \mathcal{C}_3 \}$.

The following assumptions advanced in \cite{damico2018DynamicMeasurementPoverty} define the model:\\
    ${\mathbf{A1}}$: the number $N$ of individuals in the economic system is finite and constant in time; \\
    ${\mathbf{A2}}$: the income rate processes $({\bf Y}_h)_{h \in \mathcal{H}}$ are independent and hence the class allocation processes $({\bf C}_h)_{h \in \mathcal{H}}$;\\
 %the c.a.p. $({\bf C}_h)_{h \in \mathcal{H}}$ are independent ; \\
    ${\mathbf{A3}}$: the processes $({\bf C}_h)_{h \in \mathcal{H}}$ are identically distributed ergodic Markov processes taking values in the set $\{\mathcal{C}_1,\mathcal{C}_2,\mathcal{C}_3\}$ with infinitesimal generator matrix ${\bf \Lambda}$;\\
    ${\mathbf{A4}}$: For any time $t\in \mathbb{R}$ and any individual $h\in \mathcal{H}$, the conditional distribution of the income $Y_{h}(t)$ knowing that $C_{h}(t)=\mathcal{C}_{i}$, with $\mathcal{C}_{i} \in E$, does not depend on past income values, nor on $t$ or $h$. We denote it as $F_{i}$ and we assume that it possesses finite first and second order moments. In symbol,
    \begin{equation}\label{A4for}
        {\cal{D}}(Y_{h}(t)|\sigma_{t-}(Y_{h}), C_{h}(t)=\mathcal{C}_{i})={\cal{D}}(Y_{h}(t)|C_{h}(t)=\mathcal{C}_{i})=:F_{i}(\cdot),\quad\forall t\in \mathbb{R}, \quad \forall h\in \mathcal{H},
    \end{equation}
    where $\sigma_{t-}(Y_{h}):=\lim_{s\rightarrow t^{-}}\sigma_{s}(Y_{h})$ is the sigma-algebra generated by the income process of agent $h$ up to time $t$ but excluding it.

Now, according to \cite{damico2018DynamicMeasurementPoverty} we present the stochastic extension of the poverty indexes. To this end we denote by  
\begin{equation*}
    \mathcal{P}(t)=\{h\in \mathcal{H} :  Y_{h}(t)\leq y_p\},
\end{equation*}
the set of poor agents at time $t$ and by 
\begin{eqnarray}\label{popt}
    {\mathbf{n}}(t)=\{n_1(t),n_2(t),n_3(t)\}, \quad t \in \mathbb{R}_+,
\end{eqnarray}
the multivariate counting process denoting the composition of the income classes in time. Precisely, $n_i(t)$ is the number of individuals allocated in class $\mathcal{C}_{i}$ at time $t$.

\begin{Def}
    The Dynamic Headcount ratio, The Dynamic Income-gap ratio, the Dynamic Gini and the Dynamic Sen Index are defined as follows: 
    \begin{align}
        & \mathbb{H}(t):= \frac{n_1(t)+n_2(t)}{N}, \label{H} \\
        & \mathbb{I}(t):= 1-\frac{\sum_{h\in \mathcal{P}(t)}Y_{h}(t)}{y_p (n_1(t)+n_2(t))}, \label{I} \\
        & \mathbb{G}(t):=\frac{\sum_{h\in \mathcal{P}(t)}\sum_{l \in \mathcal{P}(t)}\mid Y_{h}(t)-Y_{l}(t)\mid}{2(n_1(t)+n_2(t))(\sum_{h\in \mathcal{P}(t)} Y_{h}(t))}, \label{G} \\
        & \mathbb{S}(t)=\mathbb{H}(t)\cdot [\mathbb{I}(t)+(1-\mathbb{I}(t))\cdot \mathbb{G}(t)]. \label{S}
    \end{align}
\end{Def}

Although the previous indexes share the same functional form with their static counterparts, they are of different nature being stochastic processes due to the randomness of the counting process ${\mathbf{n}}(t)$ and of the incomes $\{Y_{h}(\cdot)\}_{h\in \mathcal{P}(t)}$.

This set of assumptions defines an economic system that describes the evolution of people according to their income. The study of this system is very complex and since the number of involved individuals $N$ is very large, it requires a big computational effort. An alternative strategy has been implemented in \cite{damico2018DynamicMeasurementPoverty} where assumption ${\mathbf{A1}}$ is relaxed in favour of a new assumption:\\
    ${\mathbf{A1'}}$: (large-size population) the number of individuals $N$ in the economy is large enough to be considered as infinity.\\
This new hypothesis allows us to use stochastic approximations based on limit theorems.

Next proposition is the first results of this strategy: 
\begin{Proposition}\label{IneqMesPropAsympIneqMeq}
    Under assumptions ${\mathbf{A1'}}$ - ${\mathbf{A4}}$, we have that:
    \begin{equation}\label{Hinf}
        \mathbb{H}(t) \stackrel{a.s.}{\longrightarrow} \mathbb{H}_{\infty}(t)=H(\mu,{\bf \Lambda},t) := \mu' \left( {\bf P}_{.1}(t)+ {\bf P}_{.2}(t)\right),
    \end{equation} 
    where $\mu'$ is the transpose of the initial distribution and ${\bf P}_{.1}(t)$ and ${\bf P}_{.2}(t)$ are the first and second column of ${\bf P}(t)=\exp(t {\bf \Lambda})$, respectively.
    
    Similarly, the Dynamic Income-gap ratio $\mathbb{I}(t)$, the Dynamic Gini index $\mathbb{G}(t)$ and the Dynamic Sen index $\mathbb{S}(t)$ converge almost surely to:
    \begin{eqnarray}\label{Largeindex}
        \mathbb{I}_{\infty}(t) & := 1 - \frac{y_1}{y_p}\frac{\mu' {\bf P}_{.1}(t)}{\mathbb{H}_{\infty}(t)} - \frac{y_2}{y_p} \frac{\mu' {\bf P}_{.2}(t)}{\mathbb{H}_{\infty}(t)}, \\
        \mathbb{G}_{\infty}(t) & := \frac{(\mu'\mathbf{P}_{.2}(t))^2\overline{z}_1+ 2(y_2-y_1)(\mu'\mathbf{P}_{.1}(t)) (\mu'\mathbf{P}_{.2}(t))+ (\mu'\mathbf{P}_{.2}(t))^2\overline{z}_2}{2\mathbb{H}_{\infty}(t)(y_1\mu'\mathbf{P}_{.1}(t)+y_2\mu'\mathbf{P}_{.2}(t))}\\ \mathbb{S}_{\infty}(t) &  := \mathbb{H}_{\infty}(t)\cdot [\mathbb{I}_{\infty}(t)+(1-\mathbb{I}_{\infty}(t))\cdot \mathbb{G}_{\infty}(t)],
    \end{eqnarray}
    where ${\bf y}=(y_{1},y_{2})$ is the vector of mean incomes per poor classes and
    \begin{equation*}
        \overline{z}_1:= \int_0^{y_{ep}}\int_0^{y_{ep}}|y-x|dF_1(y)dF_1(x),
    \end{equation*}
    \begin{equation*}
        \overline{z}_2:= \int_{y_{ep}}^{y_p}\int_{y_{ep}}^{y_p}|y-x|dF_2(y)dF_2(x).
    \end{equation*}
\end{Proposition}
\proof See Appendix.

\begin{Remark}
Proposition \ref{IneqMesPropAsympIneqMeq} was already demonstrated in \cite{damico2018DynamicMeasurementPoverty}. However, with respect to the Gini index, and in turn to the Sen index, the proof was limited to the case of equivalence of the incomes of people belonging to the same class while the hypotheses of the model advance only the equivalence of their probability distributions. The proof we provide in this paper overcomes this limitation with the addition of the inequality within each class.
\end{Remark}

The next step forward in global understanding of the time evolution of the dynamic poverty indexes is the assessment of specific central limit theorems for each index and the subsequent derivation of the confidence sets. The confidence sets are centred to the asymptotic values obtained in Proposition \ref{IneqMesPropAsympIneqMeq} and have amplitudes proportional to their variances computed in next proposition. This finding represents the main result. However, before its exposition, we anticipate an auxiliary Lemma which is a useful tool for obtaining the proof of our main result concerning the central limit theorems for the considered dynamic indexes of poverty.
\begin{Lemma}\label{lemma}
    For any $h\in \mathcal{H}$ and for any $t\in \mathbb{R}$ let $F(t;x):=\mathbb{P}[Y_h(t)\bm{\mathbbm{1}}_{\{C_h(t)\in\{C_1,C_2\}\}}\leq x]$. Then, 
    \begin{equation}
        F(t;x)= F_1(x)\mu'\mathbf{P}_{.1}(t)+F_2(x)\mu'\mathbf{P}_{.2}(t)+\mu'\mathbf{P}_{.3}(t),
    \end{equation}
    and accordingly, for any integer $r\geq 1$ 
    \begin{equation}
        \mathbb{E}[(Y_h(t)\bm{\mathbbm{1}}_{\{C_h(t)\in\{C_1,C_2\}\}})^{r}]=y_{1}^{(r)}\mu'\mathbf{P}_{.1}(t)+y_{2}^{(r)}\mu'\mathbf{P}_{.2}(t),
    \end{equation}
    where $y_{1}^{(r)}:=\mathbb{E}[(Y_1)^{r}]$ and $y_{2}^{(r)}:=\mathbb{E}[(Y_2)^{r}]$.
    
    Moreover, if $\Theta_t :=\mathbb{E}[|Y_h(t)\bm{\mathbbm{1}}_{\{C_h(t)\in\{C_1,C_2\}\}}-Y_l(t)\bm{\mathbbm{1}}_{\{C_l(t)\in\{C_1,C_2\}\}}|]$, then
    \begin{equation}\label{formtheta}
        \begin{aligned}
         	\Theta_t &=(\mu'\mathbf{P}_{.1}(t))^2\int_0^{y_{ep}}\int_0^{y_{ep}}|y-x|dF_1(x)dF_1(y) \\
            & + 2(y_2-y_1)(\mu'\mathbf{P}_{.1}(t))(\mu'\mathbf{P}_{.2}(t)) \\
            & + (\mu'\mathbf{P}_{.2}(t))^2\int_{y_{ep}}^{y_{p}}\int_{y_{ep}}^{y_{p}}|y-x|dF_2(x)dF_2(y).
        \end{aligned}
    \end{equation}
    
    Finally, using the notation
    \begin{equation}\label{eq:GINIsigma2a}
        \sigma^2(t)+\Theta^2_{t}:=\int_{-\infty}^{+\infty}\left[\int_{-\infty}^{+\infty}|y-x|dF(t;x)\right]^2dF(t;y),
    \end{equation}
    we find that
    \begin{equation}\label{eq:GINIsquaredintegral}
        \begin{aligned}
            \sigma^2(t)+\Theta^2_{t} & = (\mu'\mathbf{P}_{.1}(t))^3\int_0^{y_{ep}}\left[\int_0^{y_{ep}}|y-x|dF_1(x)\right]^2dF_1(y) \\
            & + (\mu'\mathbf{P}_{.2}(t))^2(\mu'\mathbf{P}_{.1}(t))(y_{1}^{(2)}-2y_{1}y_{2}+y_{2}^{2}) \\
            & + (\mu'\mathbf{P}_{.1}(t))^2(\mu'\mathbf{P}_{.2}(t))(y_{2}^{(2)}-2y_{1}y_{2}+y_{1}^{2})\\
            & + (\mu'\mathbf{P}_{.2}(t))^3\int_{y_{ep}}^{y_{p}}\left[\int_{y_{ep}}^{y_{p}}|y-x|dF_2(x)\right]^2dF_2(y).
         \end{aligned}
      \end{equation}
\end{Lemma} 
\proof See Appendix.

Let us introduce basic notations for expectation and variance of the random variable $Y_h(t)\bm{\mathbbm{1}}_{\{C_h(t)\in\{C_1,C_2\}\}}$:
\begin{equation}
    \overline{x}_{12}(t):= \mathbb{E}[Y_h(t)\bm{\mathbbm{1}}_{\{C_h(t)\in\{C_1,C_2\}\}}],\qquad{\sigma}_{12}^{2}(t):= \mathbb{V}[Y_h(t)\bm{\mathbbm{1}}_{\{C_h(t)\in\{C_1,C_2\}\}}]
\end{equation}

\begin{Proposition}\label{central}
    Under assumptions ${\mathbf{A1'}}$ - ${\mathbf{A4}}$, we have the following convergences in law:
    \begin{align}
        & i) \qquad \sqrt{N}\big(\mathbb{H}_N(t)-\mathbb{H}_\infty(t)\big)\xrightarrow{\mathcal{L}} \mathcal{N}\big(0,\mathbb{H}_\infty(t)(1-\mathbb{H}_\infty(t)\big);\label{cltH} \\
        & ii) \qquad\sqrt{N}\left( 1-\mathbb{I}(t)-\frac{\overline{x}_{12}(t)}{y_p\mathbb{H}_\infty(t)}\right)\xrightarrow[N\rightarrow+\infty]{\mathcal{L}}\mathcal{N}\bigg(0,\frac{\sigma_{12}^{2}(t)}{y_{p}^{2}\mathbb{H}_{\infty}^2}\bigg);\label{cltI} \\
        & iii) \qquad\sqrt{N}(\mathbb{G}_N(t)-\mathbb{G}_{\infty}(t)) \xrightarrow{\mathcal{L}} \mathcal{N}\bigg(0, \frac{\sigma^{2}(t)}{4\mathbb{H}_{\infty}^{2}(\overline{x}_{12}(t))^{2}}\bigg); \label{cltG}
    \end{align}
    where 
    \begin{equation}\label{eq:GINIsigma2}
        \sigma^2(t)=\int_{-\infty}^{+\infty}\left[\int_{-\infty}^{+\infty}|y-x|dF(t;x)\right]^2dF(t;y)-\Theta_t^2,
    \end{equation}
    and $\Theta_t$ is given in formula (\ref{formtheta}).
    \begin{align}
        & iv) \qquad\sqrt{N}(\mathbb{S}_N(t)-\mathbb{S}_\infty(t))\xrightarrow{\mathcal{L}}\mathcal{N}\Big(0,(1-\mathbb{H}_\infty(t))\frac{\mathbb{S}^2_\infty(t)}{\mathbb{H}_\infty(t)}\Big). \label{cltS}
    \end{align}
\end{Proposition} 
\proof See Appendix.

Proposition \ref{central} tells us that the dynamic poverty indexes, properly centralised and normalised, converge in distribution to normal random variables. Accordingly, it is possible to obtain their confidence sets, e.g., from formula (\ref{cltH}) we can easily determine $\forall a, b \in \mathbb{R} $ an estimate of the probability $\mathbb{P}(a\leq\mathbb{H}_N(t)\leq b)$. Indeed, once we denote by $\mathcal{Z}$ the standard normal distribution, the following approximation holds according to Proposition \ref{central}:
\begin{equation}
    \begin{aligned}
        & \mathbb{P}(a\leq\mathbb{H}_N(t)\leq b)\approx \mathbb{P}\left(\frac{a-\mathbb{H}_\infty(t)}{\sqrt{\frac{\mathbb{H}_\infty(t)(1-\mathbb{H}_\infty(t))}{N}}}\leq\mathcal{Z}\leq \frac{b-\mathbb{H}_\infty(t)}{\sqrt{\frac{\mathbb{H}_\infty(t)(1-\mathbb{H}_\infty(t))}{N}}}\right) \\
        & = \Phi\left(\frac{b-\mathbb{H}_\infty(t)}{\sqrt{\frac{\mathbb{H}_\infty(t)(1-\mathbb{H}_\infty(t))}{N}}}\right)-\Phi \left( \frac{a-\mathbb{H}_\infty(t)}{\sqrt{\frac{\mathbb{H}_\infty(t)(1-\mathbb{H}_\infty(t))}{N}}} \right).
    \end{aligned}
\end{equation}

A similar argument can be used to construct confidence sets for the other indexes. Here we just focus on the Sen index which is the most powerful index among those considered in this paper.

Let us denote by $\mathcal{E}^2(t)$ the asymptotic variance in formula (\ref{cltS}), i.e.
\begin{equation*}
    \mathcal{E}^2(t):=(1-\mathbb{H}_\infty(t))\frac{\mathbb{S}^2_\infty(t)}{\mathbb{H}_\infty(t)}.
\end{equation*}

Then, $\forall a, b \in \mathbb{R} $ the following approximation can be established:
\begin{equation}
    \begin{aligned}
         & \mathbb{P}(a\leq \mathbb{S}_N(t) \leq b) =  \mathbb{P}(a-\mathbb{S}_\infty(t))\leq \mathbb{S}_N(t)-\mathbb{S}_\infty(t)) \leq b-\mathbb{S}_\infty(t))) \\
         & = \mathbb{P}\left(\frac{a-\mathbb{S}_\infty(t))}{\frac{\mathcal{E}}{\sqrt{N}}}\leq \frac{\mathbb{S}_N(t)-\mathbb{S}_\infty(t))}{\frac{\mathcal{E}}{\sqrt{N}}}\leq \frac{b-\mathbb{S}_\infty(t))}{\frac{\mathcal{E}}{\sqrt{N}}}\right) \\
         & = \mathbb{P}\left(\frac{a-\mathbb{S}_\infty(t))}{\frac{\mathcal{E}}{\sqrt{N}}}\leq  \mathcal{Z} \leq \frac{b-\mathbb{S}_\infty(t))}{\frac{\mathcal{E}}{\sqrt{N}}}\right) \\
         & = \Phi\left(\frac{b-\mathbb{S}_\infty(t))}{\frac{\mathcal{E}}{\sqrt{N}}}\right) - \Phi\left(\frac{a-\mathbb{S}_\infty(t))}{\frac{\mathcal{E}}{\sqrt{N}}}\right).
    \end{aligned}
\end{equation}

The construction of confidence sets is of crucial importance for the application of the model since it shows the accuracy of the infinite size economic systems approximation to the real system.  

\section{Empirical application}\label{application}
    We test the model on the Italian income data provided by the Italian central bank, Banca d'Italia. The historical database is based on the survey of Italian households budgets from 1977 to 2012 and contains information about the characteristics of the individuals and their household members, along with the family net disposable incomes, which include financial assets. The poverty thresholds are reported by the Italian National Institute of Statistics (ISTAT) and are available from year 1998. However, as stated from ISTAT, the data from 1997 to 2013 are not directly comparable with the data from other years due to a substantial change in the design of the survey. Therefore, to have a clean and consistent dataset, we bound our analysis to the range 1997 to 2013. Within this range, the households data are available on a biennial basis on even years. The summary statistics for the net disposable income and grouped by number of households components are reported in Table \ref{tab:SummaryStatsY2}, while the poverty thresholds are shown in Table \ref{tab:PovertyThresholds}. The first column of both tables indicates the dimension of the household. On average, the household income increases sharply for dimensions from 1 to 3 persons, with an approximate stability from dimension 3 to 6, followed by another final increase for households bigger than 7 persons. A similar pattern is also observable to the income variability.
    
    \begin{table}
        \centering
        \resizebox{1\textwidth}{!}{
        \begin{tabular}{lrrrrrrrr}
            \hline
             &    \textbf{count} &          \textbf{mean} &            \textbf{std} &           \textbf{min} &           \textbf{25\%} &           \textbf{50\%} &           \textbf{75\%} &           \textbf{max} \\
            \hline
            1    &  14374.0 &  18233.79 &  16525.48 &  0.0 &  10795.80 &  15449.50 &  21877.92 &   810218.64 \\
            2    &  18675.0 &  29414.22 &  22868.09 &  0.0 &  17281.59 &  24662.21 &  35372.21 &   587783.94 \\
            3    &  13279.0 &  36047.38 &  24476.71 &  0.0 &  21451.87 &  31813.74 &  44322.66 &   453843.73 \\
            4    &  12089.0 &  37108.13 &  26956.98 &  0.0 &  21200.00 &  32478.33 &  46372.60 &  1022616.85 \\
            5    &   3488.0 &  35032.82 &  25323.86 &  0.0 &  18811.59 &  29715.74 &  44859.25 &   414159.70 \\
            6    &    854.0 &  37331.73 &  28507.22 &  0.0 &  19551.39 &  31298.05 &  48646.07 &   368689.73 \\
            7+   &    211.0 &  41883.71 &  46678.48 &  0.0 &  17351.48 &  31000.00 &  53253.47 &   529872.81 \\
            \hline
        \end{tabular}}
        \caption{Summary statistics of the net disposable income in euro of Italian households from 1998 to 2012 by number of family components. Data sourced from the Italian households budgets survey from Banca d'Italia.}
        \label{tab:SummaryStatsY2}
    \end{table}
    
    \begin{table}
        \centering
        \resizebox{1\textwidth}{!}{
        \begin{tabular}{lrrrrrrrr}
            \hline
             &      \textbf{1998} &      \textbf{2000} &      \textbf{2002} &      \textbf{2004} &      \textbf{2006} &      \textbf{2008} &      \textbf{2010} &      \textbf{2012} \\
            \hline
            1    &   5479.50 &   5833.54 &   5919.60 &   6623.88 &   6986.40 &   7197.60 &   7145.76 &   7134.36 \\
            2    &   9147.74 &   9722.56 &   9866.04 &  11039.76 &  11644.08 &  11996.04 &  11909.52 &  11890.56 \\
            3    &  12212.24 &  12931.00 &  13121.88 &  14682.84 &  15486.60 &  15954.72 &  15839.64 &  15814.44 \\
            4    &  14929.12 &  15847.76 &  16081.68 &  17994.84 &  18979.80 &  19553.52 &  19412.52 &  19381.56 \\
            5    &  17426.45 &  18472.86 &  18745.44 &  20975.52 &  22123.80 &  22792.44 &  22628.04 &  22592.04 \\
            6    &  19667.65 &  21000.72 &  21310.68 &  23845.92 &  25151.16 &  25911.48 &  25724.52 &  25683.60 \\
            7+   &  21963.74 &  23334.13 &  23678.52 &  26495.40 &  27945.84 &  28790.52 &  28582.80 &  28537.32 \\
            \hline
        \end{tabular}}
        \caption{Italian poverty thresholds in euro for the years 1998-2012. Data sourced from the Italian National Institute of Statistics (ISTAT)}
        \label{tab:PovertyThresholds}
    \end{table}
    
    For the application of the model with two poverty classes, we need to define an extra poverty class. We identify a class for extremely poor households by setting its threshold, $y_{ep}$, at 60\% of the poverty threshold, $y_p$. Moreover, to make the incomes comparable between the number of household components and to account for the inflation during the years, we standardise the net disposable incomes. The standardisation by the components is performed each year using the income for 1-component household as base income. Conversely, the inflation adjustment is performed setting the first year income as base income. Consequently, the poverty threshold $y_p$ is represented by the 1998 1-component value, i.e. 5479.50, and the extreme poverty threshold $y_{ep}$ is 60\% of previous threshold, i.e. 3287.70.
    
    In addition, to work on a clean sample, we require that all households show an income each year, thus we exclude households with any missing income, reducing the number of households to 914, and the number of observed incomes to 7312. Table \ref{tab:SummaryStatsByClass} reports the summary statistics for the standardised income divided by classes.
    
    \begin{table}
        \centering
        \resizebox{1\textwidth}{!}{
        \begin{tabular}{lrrrrrrrrr}
        \hline
         &   \textbf{count} & \textbf{\%}  &  \textbf{mean} &      \textbf{std} &      \textbf{min} &      \textbf{25\%} &       \textbf{50\%} &       \textbf{75\%} &        \textbf{max} \\
        \hline
        $C_1$  &   182.0 & 2.5 & 2136.62 &   977.95 &     0.00 &  1509.42 &   2354.29 &   2941.39 &    3285.68 \\
        $C_2$  &   416.0 & 5.7 & 4488.47 &   613.02 &  3293.41 &  4020.49 &   4529.34 &   5001.68   & 5467.46 \\
        $C_3$  &  6714.0 & 91.8 &  14802.99 &  9023.69 &  5480.92 &  9360.72 &  12870.52 &  17698.43 &  222822.47 \\
        \hline
        \end{tabular}}
        \caption{Summary statistics of the standardised income in euro of Italian households from 1998 to 2012 by poverty class, where $C_1$ is the class of extreme poor households and $C_2$ is the class of poor households.}
        \label{tab:SummaryStatsByClass}
    \end{table}
    
    Now, considering that we do not observe the household incomes continuously but every two years, we can estimate the generator matrix using the periodic sampling of class allocation processes described in \cite{damico2018DynamicMeasurementPoverty}. According to this methodology and with 914 independent trajectories of the class allocation processes $\bm{D}$,
    \begin{gather*}
        D_1(1998), D_1(2000),\hdots,D_1(2012), \\
        \vdots \\
        D_{914}(1998),D_{914}(2000)\hdots,D_{914}(2012),
    \end{gather*}
 where $D_{i}(t)$ denotes the income class occupied by household $i$ at year $t$.\\
\indent We first estimate the transition probability matrix $\hat{\mathbf{P}}=\hat{p}_{ij}$ with
    \begin{equation*}
        \hat{p}_{ij}=\frac{K_{ij}}{K_i},
    \end{equation*}
    where
    \begin{equation*}
        K_{ij}=\sum_{k=1}^{914}\sum_{t= 0}^{6}\mathbbm{1}_{\{D_k(2t+1998)=i,D_k(2t+2000)=j\}},
    \end{equation*}
    is the number of transitions from class $i$ to class $j$, and
    \begin{equation*}
        K_i=\sum_{j=1}^3K_{ij},
    \end{equation*}
    is the total number of times households have been allocated to class $i$.
    
    Then, the maximum likelihood estimator $\hat{\Lambda}$ of the generator matrix $\Lambda$ satisfies the relation $\hat{{\bf P}}=exp(\eta \hat{{\bf \Lambda}})$; accordingly, it can be obtained as the logarithm matrix,
    \begin{equation}
    \label{estimlamb}
        \hat{\mathbf{\Lambda}}=\frac{log(\hat{\mathbf{P}})}{\eta},
    \end{equation}
    where $\eta$ is the period of observation, i.e. 2 years in our application. It should be remarked that the maximum likelihood estimator of $\mathbf{\Lambda}$ under this observational scheme is not guaranteed to exist or to be unique (see e.g. \cite{bladt2005StatisticalInferenceDiscretely}, \cite{regnault2012EntropyEstimationQueueing}) but as proved in \cite{damico2018DynamicMeasurementPoverty}, estimator (\ref{estimlamb}) exists and is unique whenever the transition probability matrix is irreducible with positive eigenvalues. 
    
    The estimated transition probability matrix is 
    \begin{equation}\label{eq:P}
        \hat{\mathbf{P}}=
        \begin{pmatrix}
        0.37 & 0.38 & 0.25 \\
        0.11 & 0.38 & 0.51 \\
        0.01 & 0.03 & 0.96
        \end{pmatrix},
    \end{equation}
which can be readily recognised as an irreducible stochastic matrix. This matrix has positive eigenvalues:
\begin{equation*}
    x_{1}=1;\quad x_{2}=\frac{1}{200}\big(71+\sqrt{1401}\big);\quad x_{3}=\frac{1}{200}\big(71-\sqrt{1401}\big).
\end{equation*}
    Thus, the generator matrix estimated through (\ref{estimlamb}) is the  following:
    \begin{equation}\label{eq:L}
       \hat{\mathbf{\Lambda}}=
        \begin{pmatrix}
        -0.59  & 0.58 &  0.01 \\
        0.17 & -0.59  & 0.42 \\
        0.00 & 0.02 &  -0.02
        \end{pmatrix}.
    \end{equation}
    
\begin{figure}
        \subfloat{\includegraphics[width=.5\textwidth]{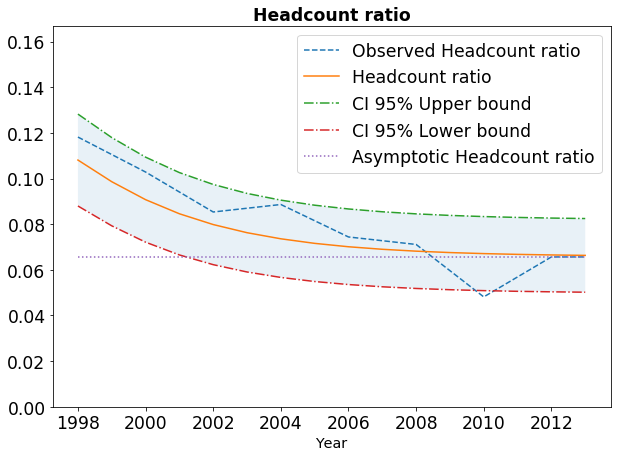}}
    	\hfill
    	\subfloat{\includegraphics[width=.49\textwidth]{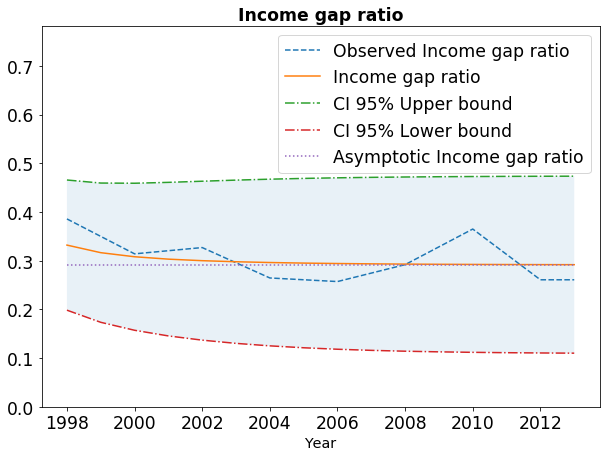}}
	
    	\subfloat{\includegraphics[width=.5\textwidth]{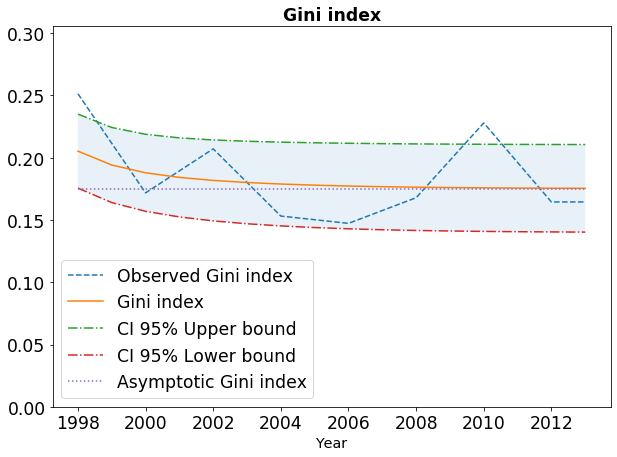}}
    	\hfill
    	\subfloat{\includegraphics[width=.5\textwidth]{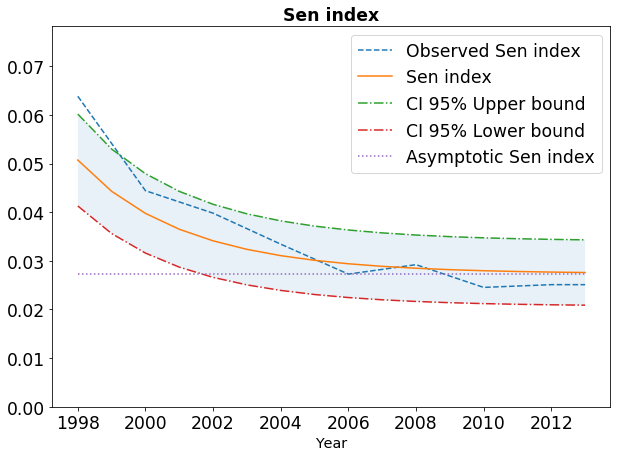}}
        \caption{Dynamic indexes of poverty for Italian households income from 1998 to 2012 computed with parameters given in (\ref{eq:P}) and (\ref{eq:mu_1998}).}
        \label{fig:plots}
    \end{figure}

    Finally, given the estimated initial distribution in the year 1998,
    \begin{equation}\label{eq:mu_1998}
        \hat{\mu}'=(0.050, 0.068, 0.882),
    \end{equation}
    and the average income for the poverty classes as reported in Table \ref{tab:SummaryStatsByClass}, we calculate the four indexes, i.e. Headcount ratio, Income gap ratio, Gini Index, and Sen index, and their respective confidence intervals at 95\% level of significance.
    Figure \ref{fig:plots} shows the four indexes estimated from the model against the observed index. In all cases, the computed indexes follow the trajectories of the observed indexes. Also, it is important to notice that the observed indexes fall within the 95\% confidence intervals with an extra small variability for the Gini index, in which the observed value in year 2010 which goes above the upper confidence limit. In general, the plots show that the model has a very good power in capturing the dynamic of the observed indexes.
    
    Figure \ref{fig:plots} also indicates that all the indexes show a decreasing path in time. This means that the different aspects of poverty represented by them are moving towards better economic conditions of the given households which include a reduction of the percentage of poor, a lower mean short-fall of people below the poverty line and a reduction of disparities among the poor. However, it is relevant to remark that at year 2012 all the indexes are very close to their stationary levels. This implies that a further decrease of poverty must necessarily be accompanied by a reinforcement of poverty containment policies or by the adoption of new ones because, if left in current conditions, the economic system cannot evolve towards a lower level of poverty. 
    
    As a robustness test for the model, we now proceed to estimate the generator matrix using a reduced set of data with only three years of observation, i.e. from 1998 to 2002. The number of households and poverty thresholds remains the same. However, the number of available incomes for the estimation reduce from 7312 to 2742. The objective of this test is to simulate a real life application in which there might be a limited history of data availability and the necessity to forecast the poverty indexes. 
    In this new setting, the estimated transition probability matrix and generator matrix become,
    \begin{equation}\label{eq:P_forecast}
        \hat{\mathbf{P}}=
        \begin{pmatrix}
        0.32 & 0.41 & 0.27 \\
        0.12 & 0.37 & 0.51 \\
        0.01 & 0.02 & 0.97
        \end{pmatrix},
        \qquad
        \hat{\mathbf{\Lambda}}=
        \begin{pmatrix}
        -0.70  & 0.69 &  0.01 \\
        0.20 & -0.63  & 0.43 \\
        0.00 & 0.02 &  -0.02
        \end{pmatrix},
    \end{equation}
    and the average income for the poverty classes become $y_1=2046.57$ and $y_2=4430.35$.
    
\begin{figure}
        \subfloat{\includegraphics[width=.5\textwidth]{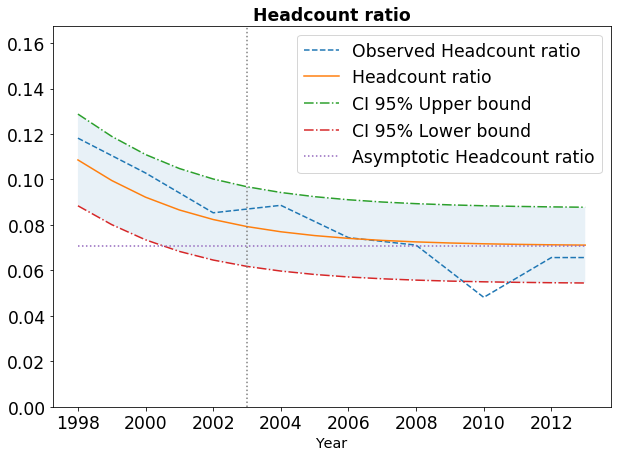}}
    	\hfill
    	\subfloat{\includegraphics[width=.49\textwidth]{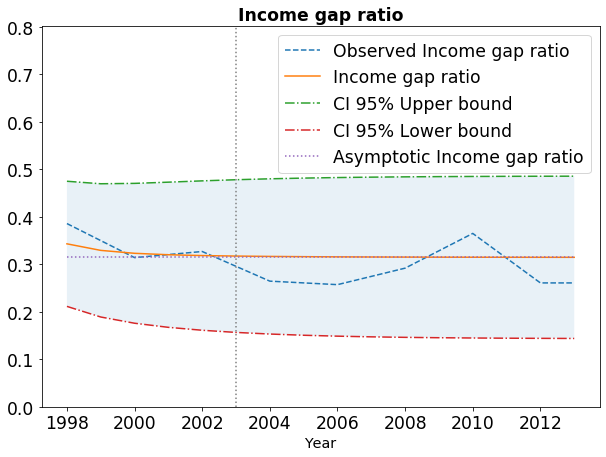}}
	
    	\subfloat{\includegraphics[width=.5\textwidth]{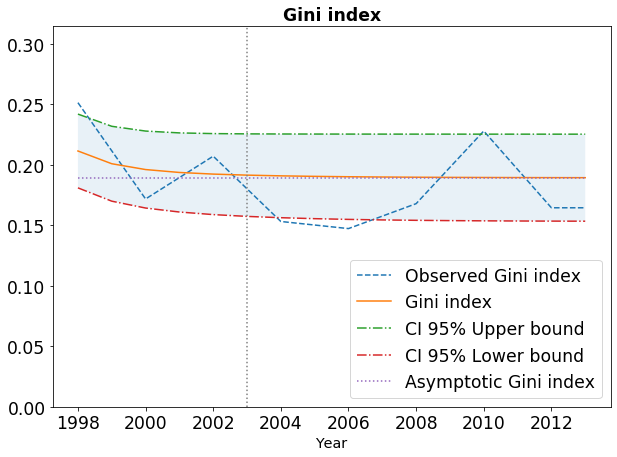}}
    	\hfill
    	\subfloat{\includegraphics[width=.5\textwidth]{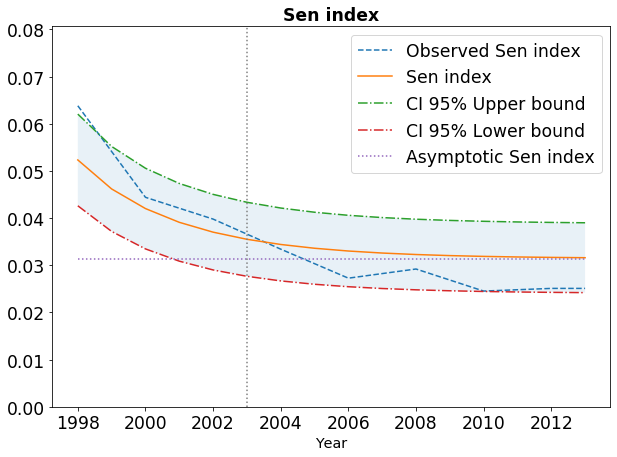}}
        \caption{Dynamic indexes of poverty for Italian households income from 1998 to 2012 computed with parameters estimated using only the initial three years of data, from 1998 to 2004. The vertical lines represent the separation between observed data on the left and forecast on the right.}
        \label{fig:plots_forecast}
    \end{figure}

    Figure \ref{fig:plots_forecast} shows that, even with a smaller dataset for the estimation procedure, the model is capable of capturing the dynamic of the observed indexes during the forecast period. This may be particularly useful when the application of the model is required to evaluate the impact of general shocks on the economic system that sometimes occur suddenly causing effects that last for several years.
    
\section{Conclusion}
    The analysis of the literature on poverty has demonstrated the importance of a dynamic approach to the determination of poverty and inequality. With the advancements proposed in this paper we aim at giving an additional tool to help the definition of the policies for the poverty in the real economies.
    
    In this study, we first proposed an extension of the dynamic Gini index, and consequently the Sen index, with the inclusion of the inequality within each class of poverty where people are classified according to their income. Then, we established the central limit theorem for each poverty index for the determination of their confidence sets. An application to the Italian income data from 1998 to 2012 confirmed the effectiveness of the considered approach and demonstrated that the model has a very good power in capturing the dynamic of the observed indexes.
    
    This study leaves some open possibilities for further research. For example, the extension to more complex dynamics or the relaxation of some of the model's assumptions. On the application side, it would be interesting to assess the model to other real economies, especially in condition of shocks, such as the recent Covid-19 disruption. 

%\section*{Bibliography}
\bibliography{CLTpoverty.bib}

\begin{thebibliography}{26}
\expandafter\ifx\csname natexlab\endcsname\relax\def\natexlab#1{#1}\fi
\providecommand{\url}[1]{\texttt{#1}}
\providecommand{\href}[2]{#2}
\providecommand{\path}[1]{#1}
\providecommand{\DOIprefix}{doi:}
\providecommand{\ArXivprefix}{arXiv:}
\providecommand{\URLprefix}{URL: }
\providecommand{\Pubmedprefix}{pmid:}
\providecommand{\doi}[1]{\href{http://dx.doi.org/#1}{\path{#1}}}
\providecommand{\Pubmed}[1]{\href{pmid:#1}{\path{#1}}}
\providecommand{\bibinfo}[2]{#2}
\ifx\xfnm\relax \def\xfnm[#1]{\unskip,\space#1}\fi
%Type = Article
\bibitem[{Annoni et~al.(2015)Annoni, Bruggemann and
  Carlsen}]{annoni2015MultidimensionalViewPoverty}
\bibinfo{author}{Annoni, P.}, \bibinfo{author}{Bruggemann, R.},
  \bibinfo{author}{Carlsen, L.}, \bibinfo{year}{2015}.
\newblock \bibinfo{title}{A multidimensional view on poverty in the {European}
  {Union} by partial order theory}.
\newblock \bibinfo{journal}{Journal of Applied Statistics}
  \bibinfo{volume}{42}, \bibinfo{pages}{535--554}.
\newblock \URLprefix \url{https://doi.org/10.1080/02664763.2014.978269},
  \DOIprefix\doi{10.1080/02664763.2014.978269}. \bibinfo{note}{publisher:
  Taylor \& Francis \_eprint: https://doi.org/10.1080/02664763.2014.978269}.
%Type = Article
\bibitem[{Bane and Ellwood(1986)}]{bane1986SlippingOutPoverty}
\bibinfo{author}{Bane, M.J.}, \bibinfo{author}{Ellwood, D.T.},
  \bibinfo{year}{1986}.
\newblock \bibinfo{title}{Slipping into and out of {Poverty}: {The} {Dynamics}
  of {Spells}}.
\newblock \bibinfo{journal}{The Journal of Human Resources}
  \bibinfo{volume}{21}, \bibinfo{pages}{1--23}.
\newblock \URLprefix \url{https://www.jstor.org/stable/145955},
  \DOIprefix\doi{10.2307/145955}. \bibinfo{note}{publisher: [University of
  Wisconsin Press, Board of Regents of the University of Wisconsin System]}.
%Type = Article
\bibitem[{Bladt and Sørensen(2005)}]{bladt2005StatisticalInferenceDiscretely}
\bibinfo{author}{Bladt, M.}, \bibinfo{author}{Sørensen, M.},
  \bibinfo{year}{2005}.
\newblock \bibinfo{title}{Statistical inference for discretely observed
  {Markov} jump processes}.
\newblock \bibinfo{journal}{Journal of the Royal Statistical Society: Series B
  (Statistical Methodology)} \bibinfo{volume}{67}, \bibinfo{pages}{395--410}.
\newblock \URLprefix
  \url{https://rss.onlinelibrary.wiley.com/doi/abs/10.1111/j.1467-9868.2005.00508.x},
  \DOIprefix\doi{10.1111/j.1467-9868.2005.00508.x}.
%Type = Article
\bibitem[{Breen and Moisio(2004)}]{breen2004PovertyDynamicsCorrected}
\bibinfo{author}{Breen, R.}, \bibinfo{author}{Moisio, P.},
  \bibinfo{year}{2004}.
\newblock \bibinfo{title}{Poverty dynamics corrected for measurement error}.
\newblock \bibinfo{journal}{The Journal of Economic Inequality}
  \bibinfo{volume}{2}, \bibinfo{pages}{171--191}.
\newblock \URLprefix \url{https://doi.org/10.1007/s10888-004-3227-9},
  \DOIprefix\doi{10.1007/s10888-004-3227-9}.
%Type = Article
\bibitem[{Cappellari and Jenkins(2004)}]{cappellari2004ModellingLowIncome}
\bibinfo{author}{Cappellari, L.}, \bibinfo{author}{Jenkins, S.P.},
  \bibinfo{year}{2004}.
\newblock \bibinfo{title}{Modelling {Low} {Income} {Transitions}}.
\newblock \bibinfo{journal}{Journal of Applied Econometrics}
  \bibinfo{volume}{19}, \bibinfo{pages}{593--610}.
\newblock \URLprefix \url{https://www.jstor.org/stable/25146308},
  \DOIprefix\doi{10.1002/jae.778}. \bibinfo{note}{publisher: Wiley}.
%Type = Article
\bibitem[{D'Amico and
  Di~Biase(2010)}]{damico2010GeneralizedConcentrationInequality}
\bibinfo{author}{D'Amico, G.}, \bibinfo{author}{Di~Biase, G.},
  \bibinfo{year}{2010}.
\newblock \bibinfo{title}{Generalized {Concentration}/{Inequality} {Indices} of
  {Economic} {Systems} {Evolving} in {Time}}.
\newblock \bibinfo{journal}{Wseas Transactions on Mathematics}
  \bibinfo{volume}{9}, \bibinfo{pages}{140--149}.
%Type = Article
\bibitem[{D'Amico et~al.(2012)D'Amico, Di~Biase and
  Manca}]{damico2012IncomeInequalityDynamic}
\bibinfo{author}{D'Amico, G.}, \bibinfo{author}{Di~Biase, G.},
  \bibinfo{author}{Manca, R.}, \bibinfo{year}{2012}.
\newblock \bibinfo{title}{Income inequality dynamic measurement of {Markov}
  models: {Application} to some {European} countries}.
\newblock \bibinfo{journal}{Economic Modelling} \bibinfo{volume}{29},
  \bibinfo{pages}{1598--1602}.
\newblock \URLprefix
  \url{http://www.sciencedirect.com/science/article/pii/S0264999312001381},
  \DOIprefix\doi{10.1016/j.econmod.2012.05.019}.
%Type = Article
\bibitem[{D'Amico et~al.(2014)D'Amico, Di~Biase and
  Manca}]{damico2014DecompositionPopulationDynamic}
\bibinfo{author}{D'Amico, G.}, \bibinfo{author}{Di~Biase, G.},
  \bibinfo{author}{Manca, R.}, \bibinfo{year}{2014}.
\newblock \bibinfo{title}{Decomposition of the {Population} {Dynamic} {Theil}'s
  {Entropy} and its application to four {European} countries}.
\newblock \bibinfo{journal}{Hitotsubashi Journal of Economics}
  \bibinfo{volume}{55}, \bibinfo{pages}{229--239}.
\newblock \URLprefix \url{https://www.jstor.org/stable/43296298}.
  \bibinfo{note}{publisher: Hitotsubashi University}.
%Type = Article
\bibitem[{D'Amico and Regnault(2018)}]{damico2018DynamicMeasurementPoverty}
\bibinfo{author}{D'Amico, G.}, \bibinfo{author}{Regnault, P.},
  \bibinfo{year}{2018}.
\newblock \bibinfo{title}{Dynamic {Measurement} of {Poverty}: {Modeling} and
  {Estimation}}.
\newblock \bibinfo{journal}{Sankhya B} \bibinfo{volume}{80},
  \bibinfo{pages}{305--340}.
\newblock \URLprefix \url{https://doi.org/10.1007/s13571-018-0153-6},
  \DOIprefix\doi{10.1007/s13571-018-0153-6}. \bibinfo{note}{citation Key Alias:
  damico2018DynamicMeasurementPovertya}.
%Type = Article
\bibitem[{Duncan et~al.(1993)Duncan, Gustafsson, Hauser, Schmauss, Messinger,
  Muffels, Nolan and Ray}]{duncan1993PovertyDynamicsEight}
\bibinfo{author}{Duncan, G.J.}, \bibinfo{author}{Gustafsson, B.},
  \bibinfo{author}{Hauser, R.}, \bibinfo{author}{Schmauss, G.},
  \bibinfo{author}{Messinger, H.}, \bibinfo{author}{Muffels, R.},
  \bibinfo{author}{Nolan, B.}, \bibinfo{author}{Ray, J.C.},
  \bibinfo{year}{1993}.
\newblock \bibinfo{title}{Poverty dynamics in eight countries}.
\newblock \bibinfo{journal}{Journal of Population Economics}
  \bibinfo{volume}{6}, \bibinfo{pages}{215--234}.
\newblock \URLprefix \url{https://doi.org/10.1007/BF00163068},
  \DOIprefix\doi{10.1007/BF00163068}.
%Type = Article
\bibitem[{Ewald and Yor(2015)}]{ewald2015IncreasingRiskInequality}
\bibinfo{author}{Ewald, C.O.}, \bibinfo{author}{Yor, M.}, \bibinfo{year}{2015}.
\newblock \bibinfo{title}{On increasing risk, inequality and poverty measures:
  {Peacocks}, lyrebirds and exotic options}.
\newblock \bibinfo{journal}{Journal of Economic Dynamics and Control}
  \bibinfo{volume}{59}, \bibinfo{pages}{22--36}.
\newblock \URLprefix
  \url{http://www.sciencedirect.com/science/article/pii/S0165188915001347},
  \DOIprefix\doi{10.1016/j.jedc.2015.07.004}.
%Type = Article
\bibitem[{Formby et~al.(2004)Formby, Smith and
  Zheng}]{formby2004MobilityMeasurementTransition}
\bibinfo{author}{Formby, J.P.}, \bibinfo{author}{Smith, W.J.},
  \bibinfo{author}{Zheng, B.}, \bibinfo{year}{2004}.
\newblock \bibinfo{title}{Mobility measurement, transition matrices and
  statistical inference}.
\newblock \bibinfo{journal}{Journal of Econometrics} \bibinfo{volume}{120},
  \bibinfo{pages}{181--205}.
\newblock \URLprefix
  \url{http://www.sciencedirect.com/science/article/pii/S0304407603002112},
  \DOIprefix\doi{10.1016/S0304-4076(03)00211-2}.
%Type = Incollection
\bibitem[{Foster(2009)}]{foster2009ClassChronicPoverty}
\bibinfo{author}{Foster, J.E.}, \bibinfo{year}{2009}.
\newblock \bibinfo{title}{A {Class} of {Chronic} {Poverty} {Measures}}, in:
  \bibinfo{editor}{Addison, T.}, \bibinfo{editor}{Hulme, D.},
  \bibinfo{editor}{Kanbur, R.} (Eds.), \bibinfo{booktitle}{Poverty {Dynamics}:
  {Interdisciplinary} {Perspectives}}. \bibinfo{publisher}{Oxford University
  Press}.
\newblock \URLprefix
  \url{https://www.oxfordscholarship.com/view/10.1093/acprof:oso/9780199557547.001.0001/acprof-9780199557547-chapter-3}.
  \bibinfo{note}{publication Title: Poverty Dynamics Section: Poverty
  Dynamics}.
%Type = Article
\bibitem[{Hirsch et~al.(2020)Hirsch, Padley, Stone and
  Valadez-Martinez}]{hirsch2020LowIncomeGap}
\bibinfo{author}{Hirsch, D.}, \bibinfo{author}{Padley, M.},
  \bibinfo{author}{Stone, J.}, \bibinfo{author}{Valadez-Martinez, L.},
  \bibinfo{year}{2020}.
\newblock \bibinfo{title}{The {Low} {Income} {Gap}: {A} {New} {Indicator}
  {Based} on a {Minimum} {Income} {Standard}}.
\newblock \bibinfo{journal}{Social Indicators Research} \bibinfo{volume}{149},
  \bibinfo{pages}{67--85}.
\newblock \URLprefix \url{https://doi.org/10.1007/s11205-019-02241-6},
  \DOIprefix\doi{10.1007/s11205-019-02241-6}.
%Type = Article
\bibitem[{Langheheine and Pol(2016)}]{langheheine2016UnifyingFrameworkMarkov}
\bibinfo{author}{Langheheine, R.}, \bibinfo{author}{Pol, F.V.D.},
  \bibinfo{year}{2016}.
\newblock \bibinfo{title}{A {Unifying} {Framework} for {Markov} {Modeling} in
  {Discrete} {Space} and {Discrete} {Time}}.
\newblock \bibinfo{journal}{Sociological Methods \& Research} \URLprefix
  \url{https://journals.sagepub.com/doi/10.1177/0049124190018004002},
  \DOIprefix\doi{10.1177/0049124190018004002}. \bibinfo{note}{publisher: SAGE
  PUBLICATIONS}.
%Type = Article
\bibitem[{Lee et~al.(2017)Lee, Ridder and
  Strauss}]{lee2017EstimationPovertyTransition}
\bibinfo{author}{Lee, N.}, \bibinfo{author}{Ridder, G.},
  \bibinfo{author}{Strauss, J.}, \bibinfo{year}{2017}.
\newblock \bibinfo{title}{Estimation of {Poverty} {Transition} {Matrices} with
  {Noisy} {Data}}.
\newblock \bibinfo{journal}{Journal of Applied Econometrics}
  \bibinfo{volume}{32}, \bibinfo{pages}{37--55}.
\newblock \URLprefix
  \url{https://onlinelibrary.wiley.com/doi/abs/10.1002/jae.2506},
  \DOIprefix\doi{10.1002/jae.2506}. \bibinfo{note}{\_eprint:
  https://onlinelibrary.wiley.com/doi/pdf/10.1002/jae.2506}.
%Type = Article
\bibitem[{Li et~al.(2001)Li, Rao and Tomkins}]{li2001LawIteratedLogarithm}
\bibinfo{author}{Li, D.}, \bibinfo{author}{Rao, M.B.},
  \bibinfo{author}{Tomkins, R.J.}, \bibinfo{year}{2001}.
\newblock \bibinfo{title}{The {Law} of the {Iterated} {Logarithm} and {Central}
  {Limit} {Theorem} for {L}-{Statistics}}.
\newblock \bibinfo{journal}{Journal of Multivariate Analysis}
  \bibinfo{volume}{78}, \bibinfo{pages}{191--217}.
\newblock \URLprefix \url{https://dl.acm.org/doi/abs/10.1006/jmva.2000.1954},
  \DOIprefix\doi{10.1006/jmva.2000.1954}.
%Type = Article
\bibitem[{McCall(1971)}]{mccall1971MarkovianModelIncome}
\bibinfo{author}{McCall, J.J.}, \bibinfo{year}{1971}.
\newblock \bibinfo{title}{A {Markovian} {Model} of {Income} {Dynamics}}.
\newblock \bibinfo{journal}{Journal of the American Statistical Association}
  \bibinfo{volume}{66}, \bibinfo{pages}{439--447}.
\newblock \URLprefix \url{https://www.jstor.org/stable/2283506},
  \DOIprefix\doi{10.2307/2283506}. \bibinfo{note}{publisher: [American
  Statistical Association, Taylor \& Francis, Ltd.]}.
%Type = Article
\bibitem[{Park and Nam(2020)}]{park2020MultidimensionalPovertyStatus}
\bibinfo{author}{Park, E.Y.}, \bibinfo{author}{Nam, S.J.},
  \bibinfo{year}{2020}.
\newblock \bibinfo{title}{Multidimensional poverty status of householders with
  disabilities in {South} {Korea}}.
\newblock \bibinfo{journal}{International Journal of Social Welfare}
  \bibinfo{volume}{29}, \bibinfo{pages}{41--50}.
\newblock \URLprefix
  \url{https://onlinelibrary.wiley.com/doi/abs/10.1111/ijsw.12401},
  \DOIprefix\doi{10.1111/ijsw.12401}. \bibinfo{note}{\_eprint:
  https://onlinelibrary.wiley.com/doi/pdf/10.1111/ijsw.12401}.
%Type = Article
\bibitem[{Parodi and Sciulli(2008)}]{parodi2008DisabilityItalianHouseholds}
\bibinfo{author}{Parodi, G.}, \bibinfo{author}{Sciulli, D.},
  \bibinfo{year}{2008}.
\newblock \bibinfo{title}{Disability in {Italian} households: income, poverty
  and labour market participation}.
\newblock \bibinfo{journal}{Applied Economics} \bibinfo{volume}{40},
  \bibinfo{pages}{2615--2630}.
\newblock \URLprefix \url{https://doi.org/10.1080/00036840600970211},
  \DOIprefix\doi{10.1080/00036840600970211}. \bibinfo{note}{publisher:
  Routledge \_eprint: https://doi.org/10.1080/00036840600970211}.
%Type = Article
\bibitem[{Regnault(2012)}]{regnault2012EntropyEstimationQueueing}
\bibinfo{author}{Regnault, P.}, \bibinfo{year}{2012}.
\newblock \bibinfo{title}{Entropy estimation for {M}/{M}/1 queueing systems}.
\newblock \bibinfo{journal}{AIP Conference Proceedings} \bibinfo{volume}{1443},
  \bibinfo{pages}{330--337}.
\newblock \URLprefix \url{https://aip.scitation.org/doi/abs/10.1063/1.3703651},
  \DOIprefix\doi{10.1063/1.3703651}. \bibinfo{note}{publisher: American
  Institute of Physics}.
%Type = Article
\bibitem[{Sen(1976)}]{sen1976PovertyOrdinalApproach}
\bibinfo{author}{Sen, A.}, \bibinfo{year}{1976}.
\newblock \bibinfo{title}{Poverty: {An} {Ordinal} {Approach} to {Measurement}}.
\newblock \bibinfo{journal}{Econometrica} \bibinfo{volume}{44},
  \bibinfo{pages}{219--231}.
\newblock \URLprefix \url{https://www.jstor.org/stable/1912718},
  \DOIprefix\doi{10.2307/1912718}. \bibinfo{note}{publisher: [Wiley,
  Econometric Society]}.
%Type = Article
\bibitem[{Shorrocks(1995)}]{shorrocks1995RevisitingSenPoverty}
\bibinfo{author}{Shorrocks, A.F.}, \bibinfo{year}{1995}.
\newblock \bibinfo{title}{Revisiting the {Sen} {Poverty} {Index}}.
\newblock \bibinfo{journal}{Econometrica} \bibinfo{volume}{63},
  \bibinfo{pages}{1225--1230}.
\newblock \URLprefix \url{https://www.jstor.org/stable/2171728},
  \DOIprefix\doi{10.2307/2171728}. \bibinfo{note}{publisher: [Wiley,
  Econometric Society]}.
%Type = Article
\bibitem[{Takayama(1979)}]{takayama1979PovertyIncomeInequality}
\bibinfo{author}{Takayama, N.}, \bibinfo{year}{1979}.
\newblock \bibinfo{title}{Poverty, {Income} {Inequality}, and {Their}
  {Measures}: {Professor} {Sen}'s {Axiomatic} {Approach} {Reconsidered}}.
\newblock \bibinfo{journal}{Econometrica} \bibinfo{volume}{47},
  \bibinfo{pages}{747--759}.
\newblock \URLprefix \url{https://www.jstor.org/stable/1910420},
  \DOIprefix\doi{10.2307/1910420}. \bibinfo{note}{publisher: [Wiley,
  Econometric Society]}.
%Type = Book
\bibitem[{Vaart(1998)}]{vaart1998AsymptoticStatistics}
\bibinfo{author}{Vaart, A.W.v.d.}, \bibinfo{year}{1998}.
\newblock \bibinfo{title}{Asymptotic {Statistics}}.
\newblock Cambridge {Series} in {Statistical} and {Probabilistic}
  {Mathematics}, \bibinfo{publisher}{Cambridge University Press},
  \bibinfo{address}{Cambridge}.
\newblock \URLprefix
  \url{https://www.cambridge.org/core/books/asymptotic-statistics/A3C7DAD3F7E66A1FA60E9C8FE132EE1D},
  \DOIprefix\doi{10.1017/CBO9780511802256}.
%Type = Article
\bibitem[{Whelan et~al.(2000)Whelan, Layte, Maître and
  Nolan}]{whelan2000POVERTYDYNAMICSAnalysis}
\bibinfo{author}{Whelan, C.T.}, \bibinfo{author}{Layte, R.},
  \bibinfo{author}{Maître, B.}, \bibinfo{author}{Nolan, B.},
  \bibinfo{year}{2000}.
\newblock \bibinfo{title}{{POVERTY} {DYNAMICS}: {An} analysis of the 1994 and
  1995 waves of the {European} {Community} {Household} {Panel} {Survey}}.
\newblock \bibinfo{journal}{European Societies} \bibinfo{volume}{2},
  \bibinfo{pages}{505--531}.
\newblock \URLprefix \url{https://doi.org/10.1080/713767003},
  \DOIprefix\doi{10.1080/713767003}. \bibinfo{note}{publisher: Routledge
  \_eprint: https://doi.org/10.1080/713767003}.

\end{thebibliography}
	
\appendix
\section{Mathematical Proofs}

{\em Proof of Proposition 2}\\
As already explained in Remark 3, we prove only the result concerning the dynamic Gini index that here is more general than that presented in \cite{damico2018DynamicMeasurementPoverty} where it is also possible to find the proof for the remaining indexes.\\
\indent From the definition of the dynamic Gini index it follows that
\begin{eqnarray}
\label{Gdiv}
\mathbb{G}(t)=\frac{\sum_{h\in \mathcal{P}(t)}\sum_{l \in \mathcal{P}(t)}\mid Y_{h}(t)-Y_{l}(t)\mid}{N^{2}}\cdot \frac{N^{2}}{2(n_1(t)+n_2(t))(\sum_{h\in \mathcal{P}(t)} Y_{h}(t))}.
\end{eqnarray}
We first determine the value to which the first factor of (\ref{Gdiv}) converges to.
Since the income classes are mutually exclusive we obtain 
\begin{equation}
    \label{4addendi}
 \begin{aligned}
     \frac{1}{N^2} \sum_{h\in\mathcal{P}(t)}\sum_{l\in\mathcal{P}(t)}|Y_h(t)-Y_l(t)| = & \frac{1}{N^2} \left\{ \sum_{h=1}^N\sum_{l=1}^N \bm{\mathbbm{1}}_{\{C_h(t)=2,C_l(t)=1\}}|Y_h(t)-Y_l(t)| \right.\\
     & + \sum_{h=1}^N\sum_{l=1}^N \bm{\mathbbm{1}}_{\{C_h(t)=1,C_l(t)=2\}}|Y_h(t)-Y_l(t)| \\
     & + \sum_{h=1}^N\sum_{l=1}^N \bm{\mathbbm{1}}_{\{C_h(t)=C_l(t)=1}\}|Y_h(t)-Y_l(t)| \\
     & \left. + \sum_{h=1}^N\sum_{l=1}^N \bm{\mathbbm{1}}_{\{C_h(t)=C_l(t)=2\}}|Y_h(t)-Y_l(t)| \right\}.
 \end{aligned}
 \end{equation}
 
 Analysing the four addends separately and observing that by construction $Y_{h}(t)>Y_{l}(t)$ for $C_{h}(t)=2$ and $C_{l}(t)=1$, we obtain:
 \begin{align*}
     & \frac{1}{N^2} \sum_{h=1}^N\sum_{l=1}^N \bm{\mathbbm{1}}_{\{C_h(t)=2,C_l(t)=1\}}|Y_h(t)-Y_l(t)| \\
     & = \frac{1}{N^2} \sum_{h=1}^N\sum_{l=1}^N \bm{\mathbbm{1}}_{\{C_h(t)=2,C_l(t)=1\}}(Y_h(t)-Y_l(t)) \\
     & = \frac{\sum_{h=1}^N\sum_{l=1}^N \bm{\mathbbm{1}}_{\{C_h(t)=2\}} \bm{\mathbbm{1}}_{\{C_l(t)=1\}} Y_h(t)- \sum_{h=1}^N\sum_{l=1}^N \bm{\mathbbm{1}}_{\{C_h(t)=2\}} \bm{\mathbbm{1}}_{\{C_l(t)=1\}} Y_l(t)}{N^2} \\
     & = \frac{\sum_{h=1}^N\bm{\mathbbm{1}}_{\{C_h(t)=2\}}Y_h(t) \sum_{l=1}^N\bm{\mathbbm{1}}_{\{C_l(t)=1\}}}{N^2}- \frac{\sum_{h=1}^N \bm{\mathbbm{1}}_{\{C_h(t)=2\}} \sum_{l=1}^N\bm{\mathbbm{1}}_{\{C_l(t)=1\}}Y_l(t)}{N^2} 
 \end{align*} 
 \begin{equation}
     \label{sost}
 = \frac{n_{C_1}(t)}{N}\cdot\frac{\sum_{h=1}^N\bm{\mathbbm{1}}_{\{C_h(t)=2\}}Y_h(t)}{N}-\frac{n_{C_2}(t)}{N}\cdot\frac{\sum_{l=1}^N\bm{\mathbbm{1}}_{\{C_l(t)=1\}}Y_l(t)}{N}
  \end{equation}
 Now observe that from the strong law of large numbers for $i\in \{1,2\}$ it holds
\begin{equation}
\label{appl1}
\frac{n_{C_1}(t)}{N} \xrightarrow{a.s.}\mathbb{E}[\mathbbm{1}_{\{C_h(t)=C_i\}}]=\mathbb{P}[C_h(t)=C_i]=\mu'\mathbf{P}_{.i}(t)
 \end{equation}
 A further application of the strong law of large numbers guarantees that
\begin{equation}
\label{appl2}
\begin{aligned}
&\frac{\sum_{l=1}^N\bm{\mathbbm{1}}_{\{C_i(t)=1\}}Y_i(t)}{N} \xrightarrow{a.s.}\mathbb{E}[\bm{\mathbbm{1}}_{\{C_i(t)=1\}}Y_i(t)]\\
& =\mathbb{E}[\bm{\mathbbm{1}}_{\{C_i(t)=1\}}\mathbb{E}[Y_i(t)|\sigma(C_{1}(s),\ldots,C_{N}(s),s\leq t)]]\\
&=\mathbb{E}[y_i\bm{\mathbbm{1}}_{\{C_i(t)=i\}}]=y_i\mathbb{P}[C_h(t)=C_i]=y_i\mu'\mathbf{P}_{.i}(t).
\end{aligned}
\end{equation}
A substitution of (\ref{appl1}) and (\ref{appl2}) in (\ref{sost}) gives
 \begin{equation}
 \begin{aligned}
 & \frac{1}{N^2} \sum_{h=1}^N\sum_{l=1}^N \bm{\mathbbm{1}}_{\{C_h(t)=2,C_l(t)=1\}}|Y_h(t)-Y_l(t)| \\
 &\xrightarrow{a.s.} \mu'\mathbf{P}_{.1}(t)y_2\mu'\mathbf{P}_{.2}(t)-(\mu'\mathbf{P}_{.2}(t))y_1\mu'\mathbf{P}_{.1}(t) = (y_2-y_1)(\mu'\mathbf{P}_{.1}(t)) (\mu'\mathbf{P}_{.2}(t)).
 \end{aligned}
 \end{equation}
  Similarly, it can be proved that
 \begin{equation*}
     \frac{1}{N^2} \sum_{h=1}^N\sum_{l=1}^N \bm{\mathbbm{1}}_{\{C_h(t)=1,C_l(t)=2\}}|Y_h(t)-Y_l(t)|  \xrightarrow{a.s.} (y_2-y_1)(\mu'\mathbf{P}_{.1}(t)) (\mu'\mathbf{P}_{.2}(t)).
 \end{equation*}
 Now, let us analyse the third addend of formula $(\ref{4addendi})$. To this end we define the random variable $Z_{h,l;1}(t)=\bm{\mathbbm{1}}_{\{C_h(t)=C_l(t)=1\}}|Y_h(t)-Y_l(t)|$ and consequently consider the following representation:
 \begin{align*}
     & \frac{1}{N^2}\sum_{h=1}^N\sum_{l=1}^N \bm{\mathbbm{1}}_{\{C_h(t)=C_l(t)=1\}}|Y_h(t)-Y_l(t)| = \frac{1}{N^2}\sum_{h\in C_1}\sum_{l\in C_1} Z_{h,l;1}(t),
 \end{align*}
 As $h$ and $l$ belong to $\mathcal{P}(t)$, we have that   $\{Z_{h,l;1}(t)\}_{h,l\in \mathcal{P}(t)}$ represents an array of $(n_{C_1}(t))^2-n_{C_1}(t)$ i.i.d. random variables with mean $\overline{z}_1$, i.e.
 \begin{equation*}
     \overline{z}_1:= \int_0^{y_{ep}}\int_0^{y_{ep}}|y-x|dF_1(y)dF_1(x).
 \end{equation*}
Again, the use of the strong law of large numbers for a random sample size gives
 \begin{align*}
     & \frac{1}{N^2}\sum_{h=1}^{n_{C_1}(t)}\sum_{l=1}^{n_{C_1}(t)} Z_{h,l;1}(t) = \frac{(n_{C_1}(t))(n_{C_1}(t)-1)}{N^2}\frac{1}{(n_{C_1}(t))(n_{C_1}(t)-1)} \sum_{h,l\in C_1} Z_{h,l;1}(t) \\
     & \xrightarrow{a.s.} (\mu'\mathbf{P}_{.1}(t))^2\cdot \overline{z}_1.
 \end{align*}
 The same reasoning can be applied to the fourth addend of (\ref{4addendi}):
 \begin{equation*}
     \frac{1}{N^2}\sum_{h=1}^{n_{C_2}(t)}\sum_{l=1}^{n_{C_2}(t)} Z_{h,l;2}(t) \xrightarrow{a.s.} (\mu'\mathbf{P}_{.2}(t))^2\cdot \overline{z}_2,
 \end{equation*}
 where
 \begin{equation*}
     \overline{z}_2:= \int_{y_{ep}}^{y_{p}}\int_{y_{ep}}^{y_{p}}|y-x|dF_2(y)dF_2(x).
 \end{equation*}
 
 Therefore, the first factor of equation (\ref{Gdiv}) converges almost surely to
 \begin{align*}
     & (\mu'\mathbf{P}_{.2}(t))^2\int_0^{y_{ep}}\int_0^{y_{ep}}|y-x|dF_1(y)dF_1(x)+ 2(y_2-y_1)(\mu'\mathbf{P}_{.1}(t)) (\mu'\mathbf{P}_{.2}(t))\\
     & + (\mu'\mathbf{P}_{.2}(t))^2\int_{y_{ep}}^{y_{p}}\int_{y_{ep}}^{y_{p}}|y-x|dF_2(y)dF_2(x).
 \end{align*}
 
 Analysing the second factor we obtain:
 \begin{align*}
     & \frac{2(n_{C_1}(t)+n_{C_2}(t))\sum_{h\in\mathcal{P}(t)}Y_h(t)}{N^2} = 2\frac{n_{C_1}(t)+n_{C_2}(t)}{N}\frac{\sum_{h\in\mathcal{P}(t)}Y_h(t)}{N}. 
 \end{align*}
Moreover we have the following convergence:
\[
2\frac{n_{C_1}(t)+n_{C_2}(t)}{N}\xrightarrow{a.s.} 2 \mathbb{E}[\bm{\mathbbm{1}}_{\{C_i(t)\in \{C_1,C_2\}\}}]=2 \mu' \left( {\bf P}_{.1}(t)+ {\bf P}_{.2}(t)\right)=
2\mathbb{H}_\infty(t),
\]
and in virtue of (\ref{appl2}) we have
\begin{equation}
\label{appl3}
\begin{aligned}
& \frac{\sum_{h\in \mathcal{P}(t)}Y_{h}(t)}{N}= \frac{\sum_{i=1}^N\bm{\mathbbm{1}}_{\{C_i(t)=C_{1}\}}Y_i(t)+\sum_{i=1}^N\bm{\mathbbm{1}}_{\{C_i(t)=C_{2}\}}Y_i(t)}{N}\\
& \xrightarrow{a.s.} y_1\mu'\mathbf{P}_{.1}(t)+y_2\mu'\mathbf{P}_{.2}(t).
\end{aligned}
\end{equation}
 Therefore, for $N\rightarrow +\infty$,
 \begin{align*}
     & \mathbb{G}_N(t) \xrightarrow{a.s.} \mathbb{G}_{\infty}(t):=\frac{(\mu'\mathbf{P}_{.2}(t))^2\overline{z}_1 + 2(y_2-y_1)(\mu'\mathbf{P}_{.1}(t)) (\mu'\mathbf{P}_{.2}(t))+ (\mu'\mathbf{P}_{.2}(t))^2\overline{z}_2}{2\mathbb{H}_{\infty}(t)(y_1\mu'\mathbf{P}_{.1}(t)+y_2\mu'\mathbf{P}_{.2}(t))}.
 \end{align*}

{\em Proof of Lemma 4}\\
In general, because the random variables $Y_i$ are non-negative, $\forall a<0$ we have $F_{i}(a)=0$ and accordingly it results
\begin{equation*}
      \begin{aligned}
 	& F(t;x):=\mathbb{P}[Y_h(t)\bm{\mathbbm{1}}_{\{C_h(t)\in\{C_1,C_2\}\}}\leq x] \\
 	&= \sum_{k=1}^3\mathbb{P}[Y_h(t)\bm{\mathbbm{1}}_{\{C_h(t)=\{C_1,C_2\}\}}\leq x | C_h(t)=C_k]\cdot \mathbb{P}[C_h(t)=C_k] \\
 	&= F_1(x)\mu'\mathbf{P}_{.1}(t)+F_2(x)\mu'\mathbf{P}_{.2}(t)+1\cdot\mu'\mathbf{P}_{.3}(t),
 \end{aligned}
\end{equation*}
where the last equality follows from assumption {{\bf A4}}. The computation of the r-th moment can be now accomplished by using the cdf $F(t;x)$. Indeed, for any integer $r\geq 1$ 
\begin{equation}
\label{integral}
\begin{aligned}
    \mathbb{E}[(Y_h(t)\bm{\mathbbm{1}}_{\{C_h(t)\in\{C_1,C_2\}\}})^{r}]&= r\int_{0}^{\infty}x^{r-1}\mathbb{P}[Y_h(t)\bm{\mathbbm{1}}_{\{C_h(t)\in\{C_1,C_2\}\}}>x]dx\\
    & =r\int_{0}^{\infty}x^{r-1}(1-F(t;x))dx.
\end{aligned}
\end{equation}
Now observe that
$
1-u'\mathbf{P}_{.3}(t)=\mu'\mathbf{P}_{.1}(t)+\mu'\mathbf{P}_{.2}(t),
$
thus
\[
1-F(t;x)=\mu'\mathbf{P}_{.1}(t)(1-F_{1}(x))+\mu'\mathbf{P}_{.2}(t)(1-F_{2}(x)).
\]
Now by substitution of the latter expression in (\ref{integral}) we have:
\begin{equation}
%\label{integral}
\begin{aligned}
    \mathbb{E}[(Y_h(t)\bm{\mathbbm{1}}_{\{C_h(t)\in\{C_1,C_2\}\}})^{r}]&= r\int_{0}^{\infty}x^{r-1}\big(\mu'\mathbf{P}_{.1}(t)(1-F_{1}(x))+\mu'\mathbf{P}_{.2}(t)(1-F_{2}(x))\big)dx\\
    & =r\mu'\mathbf{P}_{.1}(t)\int_{0}^{\infty}x^{r-1}(1-F_{1}(x))dx+r\mu'\mathbf{P}_{.2}(t)\int_{0}^{\infty}x^{r-1}(1-F_{2}(x))dx\\
    &= y_{1}^{(r)}\mu'\mathbf{P}_{.1}(t)+y_{2}^{(r)}\mu'\mathbf{P}_{.2}(t),
\end{aligned}
\end{equation}
where $y_{1}^{(r)}:=\mathbb{E}[(Y_1)^{r}]$ and $y_{2}^{(r)}:=\mathbb{E}[(Y_2)^{r}]$.\\
\indent Next point is to prove formula (\ref{formtheta}). The expected value that defines $\Theta_{t}$ can be evaluated by computing the following double integral
\[
\Theta(t)  =\int_0^{+\infty}\left(\int_0^{+\infty}|y-x|dF(x)\right)dF(y).
\]
In order to compute it, we first observe that $\forall t\in \mathbb{R}$
 \begin{equation*}
    dF(t;x)=
     \begin{cases}
    0, & \text{if } x<0\\
    dF_1(x)\mu'\mathbf{P}_{.1}(t), & \text{if } 0<x<y_{ep}\\
    dF_2(x)\mu'\mathbf{P}_{.2}(t), & \text{if } y_{ep}<x<y_{p}\\
    0, & \text{if } x>y_{p} 
    \end{cases}.
 \end{equation*}
Thus, we have 
 \begin{align*}
     \Theta(t) & = \int_0^{y_{ep}}\left(\int_0^{y_{ep}}|y-x|dF(x)\right)dF(y) + \int_{y_{ep}}^{y_{p}}\left(\int_{y_{ep}}^{y_{p}}|y-x|dF(x)\right)dF(y) \\
     & + \int_0^{y_{ep}}\left(\int_{y_{ep}}^{y_{p}}(-y+x)dF(x)\right)dF(y) + \int_{y_{ep}}^{y_{p}}\left(\int_0^{y_{ep}}(y-x)dF(x)\right)dF(y). 
 \end{align*}
 Now, we separately proceed to compute previous integrals:
 \begin{align*}
     \int_0^{y_{ep}}\left(\int_0^{y_{ep}}|y-x|dF(x)\right)dF(y) &  = \int_0^{y_{ep}}\int_0^{y_{ep}}|y-x|\mu'\mathbf{P}_{.1}(t)dF_1(x)\mu'\mathbf{P}_{.1}(t)dF_1(y) \\
     & = (\mu'\mathbf{P}_{.1}(t))^2\int_0^{y_{ep}}\int_0^{y_{ep}}|y-x|dF_1(x)dF_1(y). 
 \end{align*}
  \begin{align*}
     \int_{y_{ep}}^{y_{p}}\left(\int_{y_{ep}}^{y_{p}}|y-x|dF(x)\right)dF(y) &  = \int_{y_{ep}}^{y_{p}}\int_{y_{ep}}^{y_{p}}|y-x|\mu'\mathbf{P}_{.2}(t)dF_2(x)\mu'\mathbf{P}_{.2}(t)dF_2(y) \\
     & = (\mu'\mathbf{P}_{.2}(t))^2\int_{y_{ep}}^{y_{p}}\int_{y_{ep}}^{y_{p}}|y-x|dF_2(x)dF_2(y). 
 \end{align*}
 \begin{align*}
     & \int_0^{y_{ep}}\left(\int_{y_{ep}}^{y_{p}}(-y+x)dF(x)\right)dF(y)  = \int_0^{y_{ep}}\int_{y_{ep}}^{y_{p}}-ydF(x)dF(y)+\int_0^{y_{ep}}\int_{y_{ep}}^{y_{p}}xdF(x)dF(y)\\
     & = -\int_0^{y_{ep}}y\left(\int_{y_{ep}}^{y_{p}}\mu'\mathbf{P}_{.2}(t)dF_2(x)\right)\mu'\mathbf{P}_{.1}(t)dF_1(y)+\int_0^{y_{ep}}\left(\int_{y_{ep}}^{y_{p}}x\mu'\mathbf{P}_{.2}(t)dF_2(x)\right)\mu'\mathbf{P}_{.1}(t)dF_1(y)\\
     & = -(\mu'\mathbf{P}_{.1}(t))(\mu'\mathbf{P}_{.2}(t))\int_0^{y_{ep}}ydF_1(y)+(\mu'\mathbf{P}_{.1}(t))(\mu'\mathbf{P}_{.2}(t))\int_0^{y_{ep}}y_2dF_1(y) \\
     & = (\mu'\mathbf{P}_{.1}(t))(\mu'\mathbf{P}_{.2}(t))[-y_1+y_2]=(y_2-y_1)(\mu'\mathbf{P}_{.1}(t))(\mu'\mathbf{P}_{.2}(t)).
 \end{align*}
 
 Similarly, it is possible to prove that
 \begin{equation*}
     \int_{y_{ep}}^{y_{p}}\left(\int_0^{y_{ep}}(y-x)dF(x)\right)dF(y) =(y_2-y_1)(\mu'\mathbf{P}_{.1}(t))(\mu'\mathbf{P}_{.2}(t)).\\
 \end{equation*}
 
 Therefore, 
 \begin{equation}
 \label{eq:GINItheta}
     \begin{aligned}
        \Theta(t) & = (\mu'\mathbf{P}_{.1}(t))^2\int_0^{y_{ep}}\int_0^{y_{ep}}|y-x|dF_1(x)dF_1(y) \\
         & + 2(y_2-y_1)(\mu'\mathbf{P}_{.1}(t))(\mu'\mathbf{P}_{.2}(t)) \\
         & + (\mu'\mathbf{P}_{.2}(t))^2\int_{y_{ep}}^{y_{p}}\int_{y_{ep}}^{y_{p}}|y-x|dF_2(x)dF_2(y).
     \end{aligned}
 \end{equation}
It remains to prove formula (\ref{eq:GINIsquaredintegral}). 
For the application of this results, it remains to compute the variance $\sigma^2(t)$ in formula (\ref{eq:GINIsigma2}). In order to reach this objective we decompose the integral according to the values of $dF(t;\cdot)$. It results
 \begin{align*}
     & \int_0^{+\infty}\left[\int_0^{+\infty}|y-x|dF(t;x)\right]^2dF(t;y) \\
     & = \int_0^{y_{ep}}\left[\int_0^{y_{ep}}|y-x|dF(t;x)\right]^2dF(t;y) + \int_{y_{ep}}^{y_{p}}\left[\int_{y_{ep}}^{y_{p}}|y-x|dF(t;x)\right]^2dF(t;y) \\
     & + \int_0^{y_{ep}}\left[\int_{y_{ep}}^{y_{p}}(-y+x)dF(t;x)\right]^2dF(t;y) + \int_{y_{ep}}^{y_{p}}\left[\int_0^{y_{ep}}(y-x)dF(t;x)\right]^2dF(t;y). 
 \end{align*}
 
 Now, we separately proceed to compute previous integrals:
 \begin{align*}
     \int_0^{y_{ep}}\left[\int_0^{y_{ep}}|y-x|dF(t;x)\right]^2dF(t;y) &  = \int_0^{y_{ep}}\left[\int_0^{y_{ep}}|y-x|\mu'\mathbf{P}_{.1}(t)dF_1(x)\right]^2\mu'\mathbf{P}_{.1}(t)dF_1(y) \\
     & = (\mu'\mathbf{P}_{.1}(t))^3\int_0^{y_{ep}}\left[\int_0^{y_{ep}}|y-x|dF_1(x)\right]^2dF_1(y), 
    % & = (\mu'\mathbf{P}_{.1}(t))^3\int_0^{y_{ep}}\int_0^{y_{ep}}\int_0^{y_{ep}}|y-x||y-z|dF_1(x)dF_1(z)dF_1(y).\\
 \end{align*}
 and similarly
  \begin{align*}
     \int_{y_{ep}}^{y_{p}}\left[\int_{y_{ep}}^{y_{p}}|y-x|dF(t;x)\right]^2dF(t;y) &  = \int_{y_{ep}}^{y_{p}}\left[\int_{y_{ep}}^{y_{p}}|y-x|\mu'\mathbf{P}_{.2}(t)dF_2(x)\right]^2\mu'\mathbf{P}_{.2}(t)dF_2(y) \\
     & = (\mu'\mathbf{P}_{.2}(t))^3\int_{y_{ep}}^{y_{p}}\left[\int_{y_{ep}}^{y_{p}}|y-x|dF_2(x)\right]^2dF_2(y).
    % & = (\mu'\mathbf{P}_{.2}(t))^3\int_{y_{ep}}^{y_{p}}\int_{y_{ep}}^{y_{p}}\int_{y_{ep}}^{y_{p}}|y-x||y-z|dF_2(x)dF_2(z)dF_2(y). \\
 \end{align*}
 Furthermore we have
 \begin{align*}
     & \int_0^{y_{ep}}\left[\int_{y_{ep}}^{y_{p}}(-y+x)dF(t;x)\right]^2dF(t;y) \\
     & =\int_0^{y_{ep}}\left[\int_{y_{ep}}^{y_{p}}(-y+x)\mu'\mathbf{P}_{.2}(t)dF_2(x)\right]^2\mu'\mathbf{P}_{.1}(t)dF_1(y) \\
     & =(\mu'\mathbf{P}_{.2}(t))^{2}(\mu'\mathbf{P}_{.1}(t))\int_0^{y_{ep}}\left[-y\int_{y_{ep}}^{y_{p}}dF_2(x)+\int_{y_{ep}}^{y_{p}}xdF_2(x)\right]^2dF_1(y) \\
     & =(\mu'\mathbf{P}_{.2}(t))^{2}(\mu'\mathbf{P}_{.1}(t))\int_0^{y_{ep}}\left[-y+y_{2}\right]^2dF_1(y) \\
     & =(\mu'\mathbf{P}_{.2}(t))^{2}(\mu'\mathbf{P}_{.1}(t))\int_0^{y_{ep}}\left[y^{2}-2yy_{2}+y_{2}^{2}\right]dF_1(y) \\
     & =(\mu'\mathbf{P}_{.2}(t))^{2}(\mu'\mathbf{P}_{.1}(t))\left\{\int_0^{y_{ep}}y^{2}dF_1(y)-2y_{2}\int_0^{y_{ep}}ydF_1(y)+\int_0^{y_{ep}}y_{2}^{2}dF_1(y)\right\} \\
     & =(\mu'\mathbf{P}_{.2}(t))^{2}(\mu'\mathbf{P}_{.1}(t))\left\{y_{1}^{(2)}-2y_{1}y_{2}+y_{2}^{2}\right\}.
 \end{align*}
 
 Similarly, it is possible to prove that
 \begin{equation*}
     \int_{y_{ep}}^{y_{p}}\left[\int_0^{y_{ep}}(y-x)dF(t;x)\right]^2dF(t;y) = (\mu'\mathbf{P}_{.1}(t))^{2}(\mu'\mathbf{P}_{.2}(t))\left\{y_{2}^{(2)}-2y_{1}y_{2}+y_{1}^{2}\right\}.
 \end{equation*}
 Then by substitution we get formula (\ref{eq:GINIsquaredintegral}).\\

{\em Proof of Proposition 5}\\
{\em{i) \quad Dynamic Headcount Ratio}}\\
The random variable expressing the Headcount ratio can be expressed as a sum of i.i.d. random variables, 
\begin{equation*}
    \mathbb{H}_N(t)=\frac{n_{C_1}(t)+n_{C_2}(t)}{N}=\frac{\sum_{i=1}^N \bm{\mathbbm{1}}_{\{C_i(t)\in \{C_1,C_2\}\}}}{N},
\end{equation*}
where
\begin{equation*}
    \bm{\mathbbm{1}}_{\{C_i(t)\in \{C_1,C_2\}}=
    \begin{cases}
    1, & \mathbb{P}(C_i(t) \in \{C_1,C_2\})\\
    0, & 1-\mathbb{P}(C_i(t) \in \{C_1,C_2\}).\\
    \end{cases}
\end{equation*}
It is simple to observe that
\begin{align*}
    & \mathbb{E}[\bm{\mathbbm{1}}_{\{C_i(t)\in \{C_1,C_2\}\}}]=\mathbb{P}(C_i(t) \in \{C_1,C_2\})=\mathbb{H}_\infty(t), \\
    & V(\bm{\mathbbm{1}}_{\{C_i(t)\in \{C_1,C_2\}\}})=\mathbb{H}_\infty(t)(1-\mathbb{H}_\infty(t)).
\end{align*}
Then, as a direct application of the central limit theorem for i.i.d. random variable we can conclude that
\begin{equation*}
    \mathcal{Z}_N :=\left(\frac{\frac{\sum_{i=1}^N \bm{\mathbbm{1}}_{\{C_i(t)\in \{C_1,C_2\}\}}}{N}-\mathbb{H}(t)}{\sqrt{\frac{\mathbb{H}_\infty(t)(1-\mathbb{H}_\infty(t))}{N}}}\right)\xrightarrow[N\rightarrow +\infty]{\mathcal{L}}\mathcal{N}(0,1),
\end{equation*}
or equivalently,
\begin{equation*}
    \sqrt{N}\Bigg(\frac{\mathbb{H}_N(t)-\mathbb{H}_\infty(t)}{\sqrt{\mathbb{H}_\infty(t)(1-\mathbb{H}_\infty(t))}}\Bigg)\xrightarrow{\mathcal{L}} \mathcal{N}(0,1),
\end{equation*}
and in turn
\begin{equation*}
    \sqrt{N}\big(\mathbb{H}_N(t)-\mathbb{H}_{\infty}(t)\big)\xrightarrow{\mathcal{L}} \mathcal{N}\bigg(0,\mathbb{H}_\infty(t)(1-\mathbb{H}_\infty(t))\bigg).
\end{equation*}

{\em{ii) \quad Dynamic Income Gap Ratio}}\\
From the definition of the Income gap ratio we have that 
\begin{equation*}
    1-\mathbb{I}(t)=\frac{\sum_{h\in\mathcal{P}(t)}Y_h(t)}{y_p(n_1(t)+n_2(t))}=\frac{\frac{\sum_{h=1}^N Y_h(t)\bm{\mathbbm{1}}_{\{C_h(t)\in\{C_1,C_2\}\}}}{N}}{\frac{y_p(n_1(t)+n_2(t))}{N}}.
\end{equation*}
 Note that the denominator
\begin{equation*}
    y_p\frac{n_1(t)+n_2(t)}{N}=y_p\mathbb{H}_N(t)\xrightarrow[N\rightarrow+\infty]{a.s.}y_p\mathbb{H}_\infty(t),
\end{equation*}
as argued above in the proof of Proposition 2. The numerator expresses the sample mean of the sample $\big(\bm{\mathbbm{1}}_{\{C_1(t)\in\{C_1,C_2\}\}}Y_1(t),\ldots,\bm{\mathbbm{1}}_{\{C_N(t)\in\{C_1,C_2\}\}}Y_N(t)\big)$. According to Lemma 4, each element of this random sample has 
\begin{equation}
\label{x12}
 \mathbb{E}[Y_h(t)\bm{\mathbbm{1}}_{\{C_h(t)\in\{C_1,C_2\}\}}]=y_1\mu'\mathbf{P}_{.1}(t)+y_2\mu'\mathbf{P}_{.2}(t):=\bar{x}_{12}(t) 
\end{equation}
\begin{equation}
\begin{aligned}
	& V[Y_h(t)\bm{\mathbbm{1}}_{\{C_h(t)\in\{C_1,C_2\}\}}]=\mathbb{E}[Y^2_h(t)\bm{\mathbbm{1}}_{\{C_h(t)\in\{C_1,C_2\}\}}]-(\bar{x}_{12}(t))^2 \\
	& = [y_1^{(2)}\mu'\mathbf{P}_{.1}(t)+y_2^{(2)}\mu'\mathbf{P}_{.2}(t)]-[y_1\mu'\mathbf{P}_{.1}(t)+y_2\mu'\mathbf{P}_{.2}(t)]^2 =:\sigma_{12}^2(t).\\
\end{aligned}
\end{equation}
Then, from the CLT for i.i.d. random variable we get
\begin{equation*}
	\sqrt{N} \left( \frac{\sum_{h=1}^N Y_h(t)\bm{\mathbbm{1}}_{\{C_h(t)\in\{C_1,C_2\}\}}}{N}-\bar{x}_{12}(t)\right) \xrightarrow[N\rightarrow+\infty]{\mathcal{L}}\mathcal{Z}_{\sigma_{12}^2(t)}\sim \mathcal{N}(0,\sigma_{12}^2(t)).
\end{equation*}
Now from Slutsky's theorem (see e.g. \cite{vaart1998AsymptoticStatistics}) we can deduce that the random vector
 \begin{equation*}
 \left(
 \begin{aligned}
 	& \sqrt{N}\left( \frac{\sum_{h=1}^N Y_h(t)\bm{\mathbbm{1}}_{\{C_h(t)\in\{C_1,C_2\}\}}}{N}-\bar{x}_{12}(t)\right) \\
 	& y_p\frac{n_{C_1}(t)+n_{C_2}(t)}{N}
 \end{aligned}
 \right)
 \xrightarrow{\mathcal{L}}\left(
 \begin{aligned}
 	& \mathcal{Z}_{\sigma_{12}^2(t)} \\
 	& y_p\mathbb{H}_\infty(t)
 \end{aligned}
 \right),
 \end{equation*}
 In addition, consider the function
 $f: \mathbb{R}^{2}\rightarrow  \mathbb{R}$ defined as $f(x,y)=
 \begin{cases}
 	\frac{x}{y} & \text{ for } y\ne0 \\
 	0 & \text{ for } y=0 \\ 
 \end{cases}$, again from Slutsky's theorems we could deduce that
\begin{equation*}
 f\left(
 \begin{aligned}
 	& \sqrt{N}\left( \frac{\sum_{h=1}^N Y_h(t)\bm{\mathbbm{1}}_{\{C_h(t)\in\{C_1,C_2\}\}}}{N}-\bar{x}_{12}(t)\right) \\
 	& y_p\frac{n_{C_1}(t)+n_{C_2}(t)}{N}
 \end{aligned}
 \right)
 \end{equation*}
 \begin{equation*}
 = \frac{\sqrt{N}\left( \frac{\sum_{h=1}^N Y_h(t)\bm{\mathbbm{1}}_{\{C_h(t)\in\{C_1,C_2\}\}}}{N}-\bar{x}_{12}(t)\right)}{y_p\frac{n_{C_1}(t)+n_{C_2}(t)}{N}}
 \xrightarrow{\mathcal{L}}f\left(
 \begin{aligned}
 	& \mathcal{Z}_{\sigma_{12}^2(t)} \\
 	& y_p\mathbb{H}_\infty(t)
 \end{aligned}
 \right)=\frac{\mathcal{Z}_{\sigma_{12}^2(t)}}{y_p\mathbb{H}_\infty(t)}\sim \mathcal{N}\Big(0,\frac{\sigma_{12}^{2}(t)}{(y_p\mathbb{H}_\infty(t))^{2}}\Big)
 \end{equation*}
if $\mathbb{P}\left(\left(\begin{aligned}
 & \mathcal{Z}_{\sigma_{12}^2(t)} \\
 & y_p\mathbb{H}_\infty(t)
 \end{aligned}\right)\in C(f)\right)=1$, being $C(f)$ the continuity set of $f$. In our case,  the function $f$ is discontinuous at every point belonging to the set $\{(x,y)\in \mathbb{R}^{2}: y=0\}$  but we can observe that the probability distribution of the limiting random vector $\left(\begin{aligned}
 & \mathcal{Z}_{\sigma_{12}^2(t)} \\
 & y_p\mathbb{H}_\infty(t)
 \end{aligned}\right)$ assigns mass zero to this set, i.e.
 \[\mathbb{P}\left(\left(\begin{aligned}
 & \mathcal{Z}_{\sigma^2(t)} \\
 & y_p\mathbb{H}_\infty(t)
 \end{aligned}\right)\in \{(x,y)\in\mathbb{R}^2: \: y=0\}\right)=0,
 \]
 In this way we proved that
 \begin{equation*}
 	\left(\frac{\sqrt{N}\left( \frac{\sum_{h=1}^N Y_h(t)\bm{\mathbbm{1}}_{\{C_h(t)\in\{C_1,C_2\}\}}}{N}-\overline{x}_{12}(t)\right)}{y_p\frac{n_{C_1}(t)+n_{C_2}(t)}{N}}\right)\xrightarrow[N\rightarrow+\infty]{\mathcal{L}}\mathcal{N}\Big(0,\frac{\sigma_{12}^{2}(t)}{(y_p\mathbb{H}_\infty(t))^{2}}\Big).
 \end{equation*}
 Simple algebra and the application of the convergence  $\mathbb{H}_N(t)\xrightarrow{a.s.}\mathbb{H}_\infty(t)$ as $N\rightarrow+\infty$ produces the following result:
  \begin{equation*}
 	\sqrt{N}\left( 1-\mathbb{I}(t)-\frac{\overline{x}_{12}(t)}{y_p\mathbb{H}_\infty(t)}\right)\xrightarrow[N\rightarrow+\infty]{\mathcal{L}}\mathcal{N}\Big(0,\frac{\sigma_{12}^{2}(t)}{y_{p}^{2}\mathbb{H}_{\infty}^{2}(t)}\Big).
 \end{equation*}

% Then, for fixed $a,b\in\mathbb{R}$, we have that
 %\begin{align*}
%	 & \mathbb{P}(a\leq\mathbb{I}(t)\leq b)=\mathbb{P}(1-b\leq 1-\mathbb{I}(t)\leq 1-a)\\
%	 & =\mathbb{P}\left(\frac{(1-b)-\left[\frac{y_1\mu'\mathbf{P}_{.1}(t)+y_2\mu'\mathbf{P}_{.2}(t)}{y_p\mathbb{H}_\infty(t)}\right]}{\frac{v(t)}{\sqrt{N}}}\leq\frac{(1-\mathbb{I}(t))-\left[\frac{y_1\mu'\mathbf{P}_{.1}(t)+y_2\mu'\mathbf{P}_{.2}(t)}{y_p\mathbb{H}_\infty(t)}\right]}{\frac{v(t)}{\sqrt{N}}} \leq\frac{(1-a)-\left[\frac{y_1\mu'\mathbf{P}_{.1}(t)+y_2\mu'\mathbf{P}_{.2}(t)}{y_p\mathbb{H}_\infty(t)}\right]}{\frac{v(t)}{\sqrt{N}}}\right) \\
%	 & =\mathbb{P}\left(\frac{a-\mathbb{I}_\infty(t)}{\frac{v(t)}{\sqrt{N}}}\leq\mathcal{Z} \leq\frac{b-\mathbb{I}_\infty(t)}{\frac{v(t)}{\sqrt{N}}}\right)\\
%	 & \equiv \Phi\left(\frac{b-\mathbb{I}_\infty(t)}{\frac{v(t)}{\sqrt{N}}}\right)-\Phi\left(\frac{a-\mathbb{I}_\infty(t)}{\frac{v(t)}{\sqrt{N}}}\right).
 %\end{align*}

{\em{iii) \quad Dynamic Gini index among the poor}}
 
  We first observe that  \begin{equation*}
     \sum_{h\in\mathcal{P}(t)}\sum_{l\in\mathcal{P}(t)}|Y_h(t)-Y_l(t)|=2\sum_{1\leq l<h\leq N}|Y_h(t)\bm{\mathbbm{1}}_{\{C_h(t)\in \{C_1,C2\}\}}-Y_l(t)\bm{\mathbbm{1}}_{\{C_l(t)\in \{C_1,C2\}\}}|,
 \end{equation*}
then we represent the dynamic Gini index as
\begin{equation}
    \label{gininex}
    \mathbb{G}(t):=\frac{\frac{2\sum_{1\leq l<h\leq N}|Y_h(t)\bm{\mathbbm{1}}_{\{C_h(t)\in \{C_1,C2\}\}}-Y_l(t)\bm{\mathbbm{1}}_{\{C_l(t)\in \{C_1,C2\}\}}|}{N^{2}-N}}{\frac{2(n_1(t)+n_2(t))(\sum_{h\in \mathcal{P}(t)} Y_{h}(t))}{N^{2}-N}}.
\end{equation}
We now proceed to consider the numerator of previous formula. To the random array $(|Y_h(t)\bm{\mathbbm{1}}_{\{C_h(t)\in \{C_1,C2\}\}}-Y_l(t)\bm{\mathbbm{1}}_{\{C_l(t)\in \{C_1,C2\}\}}|)_{l,h=1}^{N}$ we can apply Theorem 3.3 in \cite{li2001LawIteratedLogarithm} and we obtain that
 \begin{align*}
 	& \sqrt{N}\left(\frac{2\sum_{h\in\mathcal{P}(t)}\sum_{l\in\mathcal{P}(t)}|Y_h(t)-Y_l(t)|}{N^2-N}-\Theta_t\right)\xrightarrow{\mathcal{L}}\mathcal{N}(0,\sigma^2(t)),
 \end{align*}
 where 
 \begin{equation}
 	\sigma^2(t)=\int_{-\infty}^{+\infty}\left[\int_{-\infty}^{+\infty}|y-x|dF(t;x)\right]^2dF(t;y)-\Theta_t^2,
 \end{equation}
 \begin{equation*}
 	\Theta_t=\mathbb{E}[|Y_h(t)\bm{\mathbbm{1}}_{\{C_h(t)\in\{C_1,C_2\}\}}-Y_l(t)\bm{\mathbbm{1}}_{\{C_l(t)\in\{C_1,C_2\}\}}|],
 \end{equation*}
 and $F(t;\cdot)$ is the cumulative distribution function of the random variable $Y_h(t)\bm{\mathbbm{1}}_{\{C_h(t)\in\{C_1,C_2\}\}}$ with $h=1,\hdots, N$, which has been derived in Lemma \ref{lemma} together with the values of $\Theta_{t}$ and $\sigma^{2}(t)$.\\
 \indent Now, consider the denominator of formula (\ref{gininex}), we have
 \begin{equation*} 
\begin{aligned}
     & \frac{2(n_{C_1}(t)+n_{C_2}(t))\sum_{h\in\mathcal{P}(t)}Y_h(t)}{N^2-N}= \frac{2}{1-\frac{1}{N}}\cdot \frac{(n_{C_1}(t)+n_{C_2}(t)}{N}\cdot \frac{\sum_{h\in\mathcal{P}(t)}Y_h(t)}{N} \xrightarrow[N\rightarrow+\infty]{{a.s.}} 2\mathbb{H}_{\infty}(t)\bar{x}_{12}(t),
 \end{aligned}
 \end{equation*}
 where the almost sure convergence is obtained by applying formulas (\ref{Hinf}), (\ref{appl3}) and (\ref{x12}). Therefore, from Slutsky's theorem, we have that
 \begin{equation}
 \label{eq:GINIconvergence}
 	\sqrt{N}\left(\frac{2\sum_{h\in\mathcal{P}(t)}\sum_{l\in\mathcal{P}(t)}|Y_h(t)-Y_l(t)|}{N^2-N}-\Theta_t\right)\cdot \frac{N^2-N}{2(n_{C_1}(t)+n_{C_2}(t))(\sum_{h\in\mathcal{P}(t)}Y_h(t))} \xrightarrow{\mathcal{L}} \mathcal{Z},
 \end{equation}
 being $\mathcal{Z}\sim\mathcal{N}(0,\mathcal{A}^2(t))$
 with $\mathcal{A}^2(t)=\frac{\sigma^2(t)}{4\mathbb{H}_{\infty}^2(t)(\bar{x}_{12}(t))^2}$.
 From equation (\ref{eq:GINIconvergence}) and from the definition of dynamic Gini index, it follows that
 \begin{equation}\label{eq:GINIconvergence2}
     \sqrt{N}\left(\mathbb{G}_N(t)-\Theta_t \cdot \frac{N^2-N}{2(n_{C_1}(t)+n_{C_2}(t))(\sum_{h\in\mathcal{P}(t)}Y_h(t))}\right) \xrightarrow{\mathcal{L}} \mathcal{N}(0,\mathcal{A}^2(t)).
 \end{equation}
 As last observation, we note that $\Theta_t$, which was evaluated in 
 (\ref{eq:GINItheta}), coincides with the numerator of $\mathbb{G}_{\infty}(t)$ and considering that
 \begin{equation*}
     \frac{N^2-N}{2(n_{C_1}(t)+n_{C_2}(t))(\sum_{h\in\mathcal{P}(t)}Y_h(t))}\xrightarrow{a.s.} \frac{1}{2\mathbb{H}_{\infty}(t)\bar{x}_{12}(t)},
 \end{equation*}
we find that equation (\ref{eq:GINIconvergence2}) is equivalent to
 \begin{equation*}
     \sqrt{N}\left(\mathbb{G}_N(t)-\Theta_t\frac{1}{2\mathbb{H}_{\infty}(t)\bar{x}_{12}(t)}\right) \xrightarrow{\mathcal{L}} \mathcal{N}(0,\mathcal{A}^2(t)),
 \end{equation*}
 and therefore
 \begin{equation*}
     \sqrt{N}(\mathbb{G}_N(t)-\mathbb{G}_{\infty}(t)) \xrightarrow{\mathcal{L}} \mathcal{N}(0,\mathcal{A}^2(t)).
 \end{equation*}
 
% This allow us to estimate $\forall a,b \in \mathbb{R}$, the probability
% \begin{align*}
%     & \mathbb{P}(a\leq \mathbb{G}_N(t) \leq b) =  \mathbb{P}(a-\mathbb{G}_\infty(t))\leq \mathbb{G}_N(t)-\mathbb{G}_\infty(t)) \leq b-\mathbb{G}_\infty(t))) \\
%     & = \mathbb{P}\left(\frac{a-\mathbb{G}_\infty(t))}{\frac{\mathcal{A}(t)}{\sqrt{N}}}\leq \frac{\mathbb{G}_N(t)-\mathbb{G}_\infty(t))}{\frac{\mathcal{A}(t)}{\sqrt{N}}}\leq \frac{b-\mathbb{G}_\infty(t))}{\frac{\mathcal{A}(t)}{\sqrt{N}}}\right) \\
%     & \equiv \Phi\left(\frac{b-\mathbb{G}_\infty(t))}{\frac{\mathcal{A}(t)}{\sqrt{N}}}\right) - \Phi\left(\frac{a-\mathbb{G}_\infty(t))}{\frac{\mathcal{A}(t)}{\sqrt{N}}}\right).
 %\end{align*}

{\em{iv) \quad Dynamic Sen Index}}

The Dynamic Sen Index is defined according to $  \mathbb{S}(t)=\mathbb{H}(t)[\mathbb{I}(t)+(1-\mathbb{I}(t))\mathbb{G}(t)]$. Since we proved that $\mathbb{I}(t)\xrightarrow{a.s.}\mathbb{I}_\infty(t)$ and $\mathbb{G}(t)\xrightarrow{a.s.}\mathbb{G}_\infty(t)$, then from the continuous mapping theorem (see, e.g. \cite{vaart1998AsymptoticStatistics}) we can write 
 \begin{equation*}
    [\mathbb{I}(t)+(1-\mathbb{I}(t))\mathbb{G}(t)]\xrightarrow{a.s.}[\mathbb{I}_\infty(t)+(1-\mathbb{I}_\infty(t))\mathbb{G}_\infty(t)]\in \mathbb{R}.
 \end{equation*}
 Furthermore, from point $i)$ of Proposition \ref{central} we know that
 \begin{equation*}
     \sqrt{N}(\mathbb{H}_N(t)-\mathbb{H}_\infty(t))\xrightarrow[N\rightarrow+\infty]{\mathcal{L}}\mathcal{N}\big(0,\mathbb{H}_\infty(t)(1-\mathbb{H}_\infty(t))\big),
 \end{equation*}
and then  if we consider the function $f: \mathbb{R}^{2}\rightarrow  \mathbb{R}$ defined as $f(x,y)=xy$ due to the continuity of $f$ we have
 \begin{equation*}
     f\left(
     \begin{aligned}
        & \sqrt{N}(\mathbb{H}_N(t)-\mathbb{H}_\infty(t)) \\
        & \mathbb{I}(t)+(1-\mathbb{I}(t))\mathbb{G}(t) \\
     \end{aligned}
     \right)
     \xrightarrow{\mathcal{L}}f\left(
     \begin{aligned}
        & \mathcal{Z}_{\sigma^2_\mathbb{H}(t)} \\
        & \mathbb{I}_\infty(t)+(1-\mathbb{I}_\infty(t))\mathbb{G}_\infty(t) \\
     \end{aligned}
     \right)=\mathcal{Z}_{\sigma^2_\mathbb{H}(t)}\cdot \big(\mathbb{I}_\infty(t)+(1-\mathbb{I}_\infty(t))\mathbb{G}_\infty(t)\big),
 \end{equation*}
 where $\mathcal{Z}_{\sigma^2_\mathbb{H}(t)}\sim\mathcal{N}(0,\mathbb{H}_\infty(t)(1-\mathbb{H}_\infty(t))$.
 
 Thus, we have that 
 \begin{equation*}
     \sqrt{N}(\mathbb{H}_N(t)-\mathbb{H}_\infty(t))[\mathbb{I}_N(t)+(1-\mathbb{I}_N(t))\mathbb{G}_N(t)]\xrightarrow[N\rightarrow+\infty]{\mathcal{L}}\mathcal{N}\Big(0,\sigma^2_\mathbb{H}(t)(\mathbb{I}_\infty(t)+(1-\mathbb{I}_\infty(t))\mathbb{G}_\infty(t))^2\Big)
 \end{equation*}
 Simple algebraic manipulations give  \begin{equation*}
     \sqrt{N}(\mathbb{S}_N(t)-\mathbb{H}_\infty(t)[\mathbb{I}_N(t)+(1-\mathbb{I}_N(t))\mathbb{G}_N(t)])
     \xrightarrow{\mathcal{L}} \mathcal{N}\Big(0,(1-\mathbb{H}_\infty(t))\frac{\mathbb{S}^2_\infty(t)}{\mathbb{H}_\infty(t)}\Big),
 \end{equation*}
 but since $[\mathbb{I}_N(t)+(1-\mathbb{I}_N(t))\mathbb{G}_N(t)]\xrightarrow{a.s.}[\mathbb{I}_\infty(t)+(1-\mathbb{I}_\infty(t))\mathbb{G}_\infty(t)]$, we can conclude that
 \begin{equation*}
     \sqrt{N}(\mathbb{S}_N(t)-\mathbb{S}_\infty(t))\xrightarrow{\mathcal{L}}\mathcal{N}\Big(0,(1-\mathbb{H}_\infty(t))\frac{\mathbb{S}^2_\infty(t)}{\mathbb{H}_\infty(t)}\Big).
 \end{equation*}

\end{document}